\documentclass[twocol]{ametsocV6.1}
\usepackage{enumitem}

\title{Origin and Limits of Invariant Warming Patterns in Climate Models}

\authors{Paolo Giani,\aff{a}\correspondingauthor{Paolo Giani, pgiani@mit.edu} Arlene M. Fiore,\aff{a,b} Glenn Flierl,\aff{a}  Raffaele Ferrari\aff{a}, Noelle E. Selin\aff{a,b,c} }

\affiliation{\aff{a}{Massachusetts Institute of Technology, Earth, Atmospheric, and Planetary Sciences, Cambridge, MA, USA}\\ \aff{b}{Massachusetts Institute of Technology, Center for Sustainability, Science and Strategy, Cambridge, MA, USA} \\ \aff{c}{Massachusetts Institute of Technology, Institute for Data, Systems, and Society, Cambridge, MA, USA}}

\abstract{Climate models exhibit an approximately invariant surface warming pattern in typical end-of-century projections. This observation has been used extensively in climate impact assessments for fast calculations of local temperature anomalies, with a linear procedure known as \emph{pattern scaling}. At the same time, emerging research has also shown that time-varying warming patterns are necessary to explain the time evolution of effective climate sensitivity in coupled models, a mechanism that is known as the \emph{pattern effect} and that seemingly challenges the \emph{pattern scaling} understanding. Here we present a simple theory based on local energy balance arguments to reconcile this apparent contradiction. Specifically, we show that the pattern invariance is an inherent feature of exponential forcing, linear feedbacks, a constant forcing pattern and diffusive dynamics. These conditions are approximately met in most CMIP6 Shared Socioeconomic Pathways (SSP), except in the Arctic where nonlinear feedbacks are important and in regions where aerosols considerably alter the forcing pattern. In idealized experiments where concentrations of CO$_2$ are abruptly increased, such as those used to study the \emph{pattern effect}, the warming pattern can change considerably over time because of spatially inhomogeneous ocean heat uptake, even in the absence of nonlinear feedbacks. Our results illustrate why typical future projections are amenable to pattern scaling, and provide a plausible explanation of why more complicated approaches, such as nonlinear emulators, have only shown marginal improvements in accuracy over simple linear calculations.}

\begin{document}

\maketitle

%
%
%
\statement
In typical end-of-century climate projections from comprehensive models, the ratio between local and global surface temperature anomalies is approximately time- and scenario-invariant. This observation has enabled fast calculations of local temperature changes by scaling the global average with a constant pattern. At the same time, idealized 4xCO2 experiments show a different behavior and a considerable time-evolution of the warming pattern. We present a simple theory based on local energy balance to reconcile this apparent contradiction. Specifically, we show that the pattern invariance arises under a set of conditions that are approximately satisfied in end-of-century scenarios but not in abrupt idealized experiments.  Our findings clarify why scaling the global average to calculate local temperature anomalies is effective for most future projections.
%
%


\section{Introduction}

In typical end-of-century projections, climate models show an approximately linear relationship between local and global surface air temperature anomalies \citep{tebaldi_pattern_2014,osborn_performance_2018}. Mathematically, this implies that the warming pattern $\Delta T(\mathbf{r},t)/\overline{\Delta T}(t)$ is approximately independent of both time and the forcing magnitude, where $\Delta T(\mathbf{r},t)$ is the surface air temperature response at location $\mathbf{r}$ and time $t$, and the overbar denotes spatial averaging. This diagnostic result from climate models has been recognized since the early work of \citet{santer_developing_1990} and it underlies the so-called pattern scaling approach upon which fast emulators of climate models are built \citep{mitchell_pattern_2003,beusch_emulating_2020}. More recently, \cite{leduc_regional_2016} noted that regional temperature anomalies are also linearly related to cumulative carbon emissions, and defined the Regional Transient Climate Response to cumulative carbon Emissions ($\text{RTCRE} = \Delta T(\mathbf{r},t)/\sum_tE(t)$) that quantifies the amount of regional warming from total emitted carbon $\sum_tE(t)$. The RTCRE property underlies policy-relevant calculations of carbon emissions impacts on the global economy and the biosphere \citep{estrada_global_2017,hsiang_estimating_2017,stewart_climate_2020}, as well as national attributions of historical climate damages \citep{callahan_national_2022}. 

The near-invariance of both the RTCRE and the warming pattern $\Delta T(\mathbf{r},t)/\overline{\Delta T}(t)$ implies that the ratio $\overline{\Delta T}(t)/\sum_t E(t)$ is also constant on a centennial time scale \citep{raupach_exponential_2013,zickfeld_is_2012}. This ratio is known as the Transient Climate Response to Cumulative CO$_2$ Emissions (TCRE), and is also recognized to be approximately invariant in climate models \citep{matthews_proportionality_2009}. The recognition of a constant TCRE shaped the concept of carbon budgets to meet global temperature goals set by international agreements \citep{matthews_opportunities_2020,matthews_stabilizing_2008}. The TCRE is qualitatively understood as a compensation effect between the decreasing radiative forcing per unit mass of CO$_2$ at increasing CO$_2$ levels and the decreasing efficiency of ocean heat and carbon uptake in a warmer planet \citep{solomon_irreversible_2009,macdougall_transient_2016,macdougall_oceanic_2017}. 
While the TCRE property has been the focus of many studies \citep{bronselaer_heat_2020,gillett_warming_2023,seshadri_origin_2017}, we lack a comprehensive theory on why climate models show an approximately invariant $\Delta T(\mathbf{r},t)/\overline{\Delta T}(t)$ and RTCRE, despite their practical importance.

Furthermore, recent research appears to contradict the basis for pattern scaling by demonstrating that a non-constant $\Delta T(\mathbf{r},t)/\overline{\Delta T}(t)$ is required to explain time-varying effective climate sensitivity in climate models \citep{armour_time-varying_2013,andrews_dependence_2015}, since radiative feedbacks vary considerably in space and evolving warming patterns "activate" feedbacks of different strengths. The dependency of the effective climate sensitivity on the evolving warming patterns has been termed the \emph{pattern effect} \citep{rugenstein_dependence_2016,andrews_accounting_2018}, and has far-reaching implications for interpreting climate sensitivity estimates from historical observations \citep{armour_energy_2017,armour_sea-surface_2024}. Evolving patterns of surface warming can be due to both unforced coupled atmosphere-ocean variability on annual and decadal timescales, as well as  forced responses on decadal to centennial timescales \citep{andrews_accounting_2018,armour_energy_2017,knutti_beyond_2017}.

In this work, we aim to reconcile the apparent contradiction between \emph{pattern scaling} and the \emph{pattern effect} and to provide a simple theory that explains why we observe a rather constant $\Delta T(\mathbf{r},t)/\overline{\Delta T}(t)$ in end-of-century projections. We use a combination of local energy balance arguments and model data to argue that this is inherent to the nature of exponential forcing, linear feedbacks, an invariant forcing pattern and diffusive dynamics. We discuss regions and scenarios where this approximation fails, notably the Arctic and locations where aerosol forcing differs across scenarios. In idealized experiments where CO$_2$ is abruptly increased in the atmosphere, the warming pattern evolves because of the transient adjustment to a new equilibrium.

This paper is organized as follows. Section \ref{sec:CMIP6Evidence} illustrates the concepts of pattern scaling and the pattern effect by presenting diagnostics from state-of-the-art climate models. Section \ref{sec:idealized} introduces the local energy balance and several idealizations that we use to explain the diagnostic results of Section \ref{sec:CMIP6Evidence}. Section \ref{sec:interpretation} offers interpretations of the idealized results in the context of coupled models. Section \ref{sec:conclusions} summarizes the main findings and discusses the main implications of our results beyond this work.

\section{Warming patterns in CMIP6 across time and experiments}

\label{sec:CMIP6Evidence}

We consider data from the latest iteration of the Coupled Model Intercomparison Project (CMIP6). We examine the models' responses to substantially different forcing from two Shared Socioeconomic Pathways (ssp119 and ssp585) and an idealized experiment from CMIP6 DECK \citep{eyring_overview_2016} where CO$_2$ levels are abruptly quadrupled (abrupt-4xCO2). The two future scenarios (ssp119 and ssp585) are the lowest and highest emissions scenarios considered in ScenarioMIP, respectively \citep{oneill_scenario_2016}. We analyze the \emph{forced} response by computing the CMIP6 multi-model average in the three scenarios, after regridding all the model output to a common 2 degree resolution grid. We use all models in CMIP6 that ran ssp119, ssp585, abrupt-4xCO2 and the preindustrial control (piControl) experiments.

\begin{figure*}[t]
\centerline{\includegraphics[width=\textwidth]{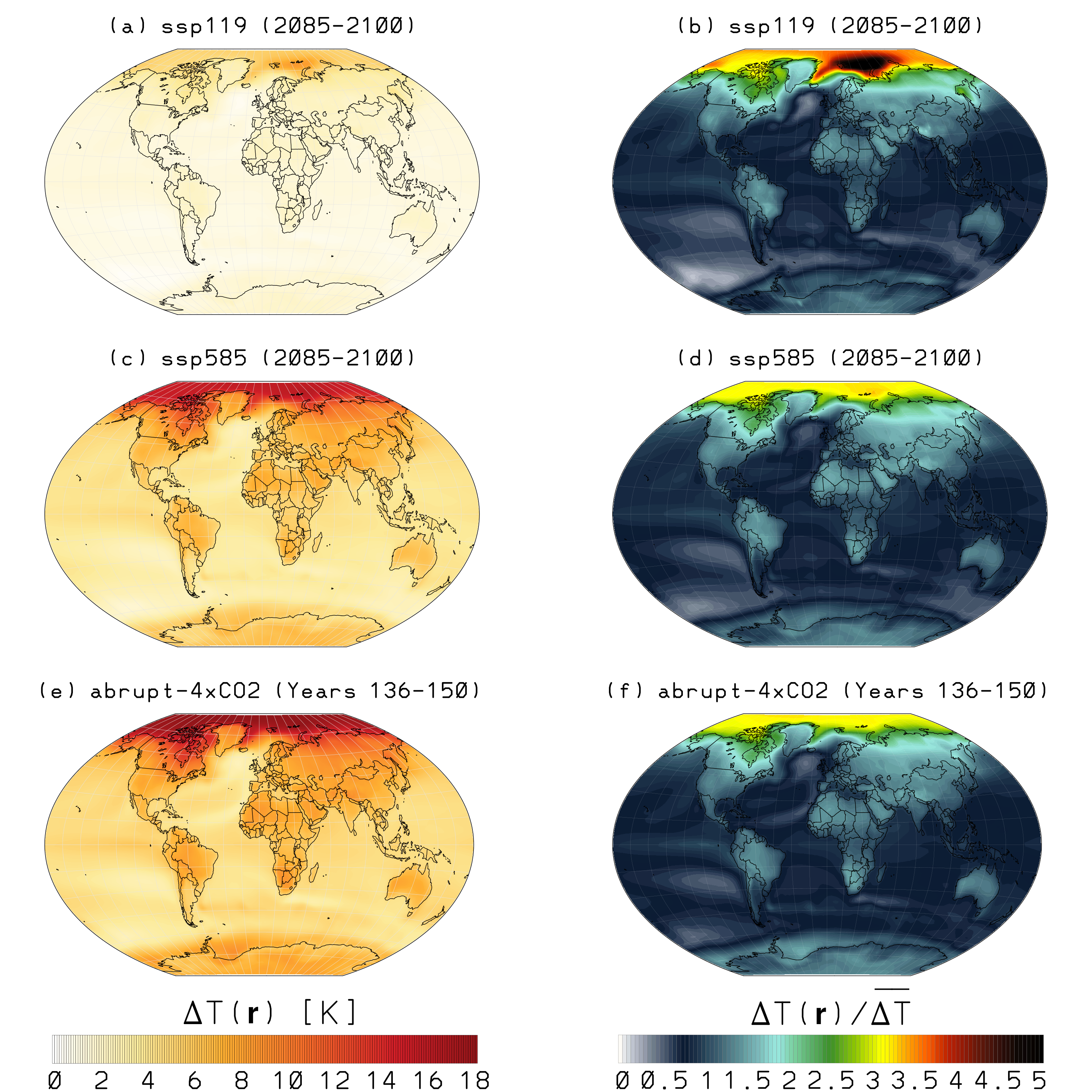}}
  \caption{CMIP6 multi-model averages of temperature anomalies (a,c,e) and warming patterns (b,d,f). Data are averaged over the final 15 years of ssp119 (top row), ssp585 (middle row) and abrupt-4xCO2 (bottom row).}
  \label{fig:cmipPattern}
\end{figure*}

Figure \ref{fig:cmipPattern} shows the end-of-century surface regional warming $\Delta T(\mathbf{r})$ in the two different SSPs projections and abrupt-4xCO2 (averaged over years 136--150 after the initial CO$_2$ quadrupling), along with the regional warming normalized by the respective global average (i.e., the pattern $\Delta T(\mathbf{r})/\overline{\Delta T}$). Throughout this work, $\Delta$ refers to changes with respect to preindustrial climatology. Except for a few regions, regional warming collapses on a single pattern when normalized by the respective global average.
This is the basis for pattern scaling, and is quite remarkable given the extremely different forcing in ssp126, ssp585 and abrupt-4xCO2. Moreover, the pattern remains strikingly similar for other averaging periods and future scenarios. Specifically, we observe a similar pattern of enhanced warming ($\Delta T(\mathbf{r})/\overline{\Delta T} > 1$) over land and the Arctic region, with less-than-average warming ($\Delta T(\mathbf{r})/\overline{\Delta T} < 1$) over the ocean, with exceptionally small warming over the North Atlantic warming hole and the Southern Ocean. These are rather well-known fingerprints of global warming predicted by coupled models, and several studies have investigated the physical reasons underlying these trends. Atmospheric temperatures over land are expected to warm more than over the ocean because of a combination of lower thermal inertia inland (which matters during the transient adjustment to equilibrium), lower moisture availability for evaporation and a smaller decrease in the lapse rate under warming, among other factors \citep{byrne_trends_2018,sutton_landsea_2007,joshi_mechanisms_2008}. Many studies have also focused on the mechanisms related to enhanced Arctic warming, and linked it to the sea-ice albedo and insulation feedbacks, the positive lapse-rate feedback and the poleward increase in atmospheric moisture transport \citep{previdi_arctic_2021,serreze_processes_2011}. The North Atlantic Warming Hole is understood to be the result of ocean processes, including lower heat import from the slowdown of the meridional overturning circulation in the Atlantic and increased ocean heat export into the Arctic, as well as atmospheric processes related to cloud feedbacks and aerosols \citep{keil_multiple_2020}. In stark contrast with the Arctic, the Southern Ocean is projected to warm less than the global average in the SSPs as a result of strong wind-driven upwelling of unchanged water from depth and anomalous equatorward heat transport by climatological ocean currents \citep{armour_southern_2016}.

\begin{figure*}[t]
\centerline{\includegraphics[width=\textwidth]{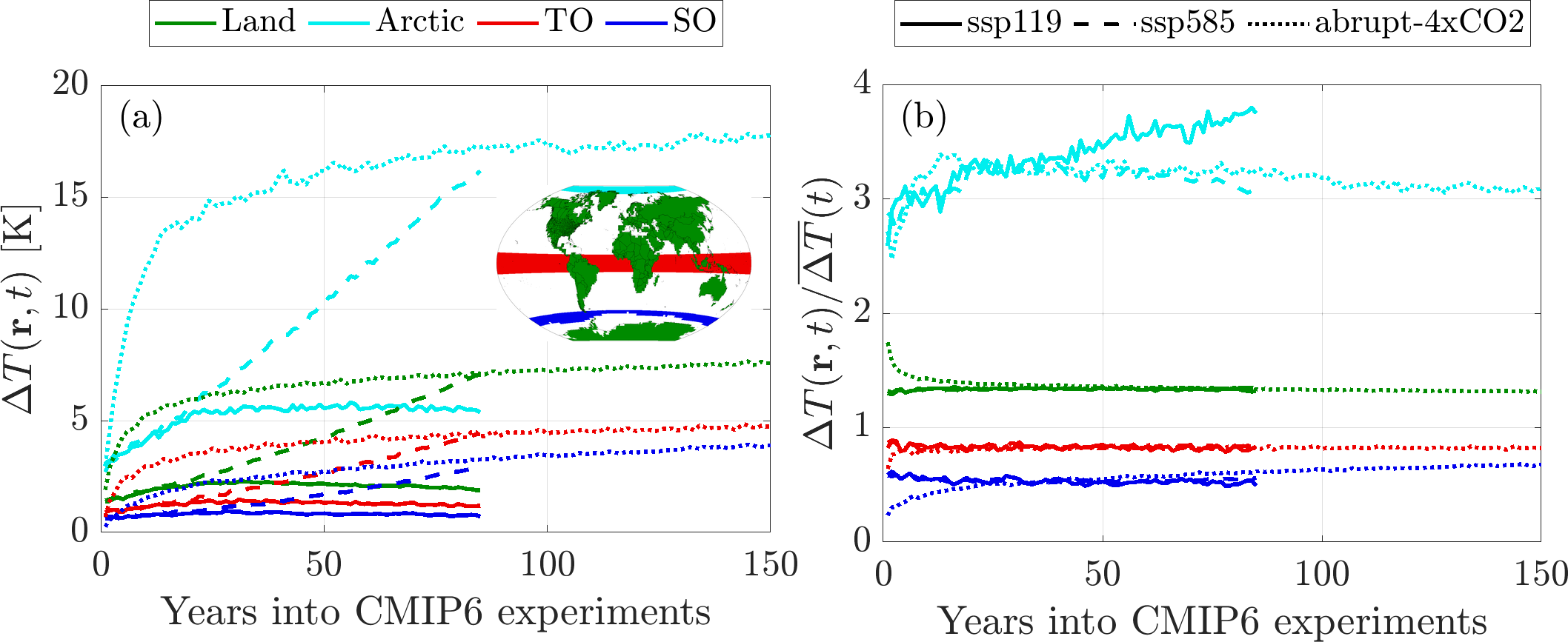}}
  \caption{CMIP6 multi-model averages of temperature anomalies (a) and warming patterns (b) averaged over the four regions described in the main text (TO=Tropical Ocean, SO=Southern Ocean). The inset in the left panel shows the averaging regions.}
  \label{fig:RegionalPatterns}
\end{figure*}

Despite the striking similarities of the warming pattern across emission scenarios, there are regions (such as the Arctic) where the pattern is different across time and scenarios. To further investigate and elucidate this dependence, we select four regions for additional study.  Specifically, in Figure \ref{fig:RegionalPatterns} we average all grid points located over \emph{Land}, over the high-latitude \emph{Arctic} region (defined as poleward of 80$^\circ$N latitude), over the tropical ocean (\emph{TO}, defined as ocean points within $\pm$10$^\circ$ latitude) and over the portion of the Southern Ocean that is relatively ice-free (\emph{SO}, defined as ocean points poleward of 55$^\circ$S and with annually-averaged sea ice concentration lower than 30\% during the preindustrial control). The four different regions warm at different rates in each scenario (Figure \ref{fig:RegionalPatterns}a), but the two  SSPs collapse on a single constant line over Land, TO and SO when normalized by the global average. In abrupt-4xCO2, the warming over Land and TO also collapses to the same constant line after some initial adjustment, whereas over SO the pattern varies over time. This is linked to the \emph{pattern effect} discussed in more detail in Section \ref{sec:idealized} and Section \ref{sec:interpretation}. Over the Arctic region, the three experiments have distinctive patterns that also change in time, although the average warming remains constrained between 3 and 4 times the global average (i.e., it exhibits relatively limited, but clearly distinguishable, changes across time and scenarios). At a local scale (i.e., without averaging across regions), the behavior is similar but some grid points display larger variations, especially in the Arctic (Figure \ref{fig:cmipPattern}).

These calculations are in line with previous studies \citep{leduc_regional_2016,wells_understanding_2023}, and offer a diagnostic perspective on why many climate impact assessments use pattern scaling to compute regional warming. However, they do not necessarily provide insights into the mechanisms for the relative invariance of $\Delta T(\mathbf{r},t)/\overline{\Delta T}(t)$. What determines the pattern that seems to be consistent across time and SSPs? Why is the pattern scenario- and time-independent in the SSPs, except for a few regions? We explore these questions with several idealizations based on local energy balance (Section \ref{sec:idealized}), and discuss in the context of CMIP6 data and previous studies (Section \ref{sec:interpretation}).

\section{Warming Patterns from Idealized Local Energy Balance}

In this Section, we first introduce the local energy balance framework and a set of four assumptions that are sufficient to make the warming pattern scenario- and time-independent. We then relax each assumption individually or in combination with others,  to disentangle the different ways that the warming pattern can change.

\label{sec:idealized}

\subsection{Local Energy Balance}

\label{subsec:localbalance}

The top-of-atmosphere (TOA) local energy balance  provides a useful conceptual tool to study the time evolution (or lack thereof) of $\Delta T(\mathbf{r},t)/\overline{\Delta T}(t)$. The energy content of a generic atmosphere-ocean column at a geographic location $\mathbf{r} = \left( \phi, \theta \right) = $ (latitude, longitude) responds to changes in local radiative properties and heat divergence:
\begin{equation}
H(\mathbf{r},t) = \Delta N(\mathbf{r},t) + \Delta \left( \nabla \cdot \mathbf{F}(\mathbf{r},t)\right)
\label{eq:localbalance}
\end{equation}
Where $\Delta$ represents a perturbation with respect to a previous state (preindustrial control), $H(\mathbf{r},t)$ is the local rate of change of energy content in the column, $N(\mathbf{r},t)$ is the local net energy imbalance at the top of the atmosphere and $\nabla \cdot \mathbf{F}(\mathbf{r},t)$ is the horizontal heat flux divergence.

The general form of Equation \ref{eq:localbalance} does not immediately provide quantitative information about $\Delta T(\mathbf{r},t)/\overline{\Delta T}(t)$. 
We proceed by making four simplifications to derive a version of Equation (\ref{eq:localbalance}) where the ratio $\Delta T(\mathbf{r},t)/\overline{\Delta T}(t)$ is perfectly time- and scenario-independent.  We stress that the following assumptions are not necessarily realistic; however, they serve as a useful starting point to understand the mechanisms of evolving warming patterns. We will test the extent to which these simplifications are valid in CMIP6 experiments in Section \ref{sec:interpretation}. The four conditions that, together, are sufficient to make $\Delta T(\mathbf{r},t)/\overline{\Delta T}(t)$ scenario- and time-independent are the following:

\begin{enumerate}[label=(\roman*),ref=\roman*]

    \item The system is in equilibrium, i.e. $H(\mathbf{r},t) = 0$.

\item The energy balance at the top of the atmosphere is described with a local linearization with time-invariant and stabilizing feedback parameters \citep{feldl_nonlinear_2013,pfister_changes_2021,armour_time-varying_2013}:
    \begin{equation}
    \Delta N(\mathbf{r},t) = R(\mathbf{r},t) + \lambda(\mathbf{r})\Delta T(\mathbf{r},t)
    \label{eq:linearimbalance}
    \end{equation}
    Where $R$ is the effective radiative forcing, $\lambda(\mathbf{r})$ is the \emph{local} and time-invariant feedback parameter, and $\Delta T(\mathbf{r},t)$ is the \emph{local} temperature change. The first term $R(\mathbf{r},t)$ is the energy imbalance generated by some forcing agent including all the atmospheric adjustments with fixed surface temperature. The second term $\lambda(\mathbf{r})\Delta T(\mathbf{r},t)$ is the radiative response, which is assumed to be proportional to surface temperature $\Delta T(\mathbf{r},t)$ through the feedback parameter $\lambda(\mathbf{r})$. In our notation, a negative value of $\lambda(\mathbf{r})$ is stabilizing. In Appendix A, we show that (ii) could be relaxed to also include the \emph{nonlocal} dependence of $\Delta N$ on surface temperatures at all locations.

    \item The forcing function is described as the product of the globally-averaged radiative forcing $\overline{R}(t)$ and a time-invariant dimensionless constant forcing pattern $P_R(\mathbf{r})$. In other words, the time- and space-dependence of the forcing function can be separated into $R(\mathbf{r},t) = P_R(\mathbf{r})\overline{R}(t)$. By definition, $\overline{P_R}(\mathbf{r}) = 1$.

    \item Changes in horizontal heat flux divergence are negligible, i.e. $\Delta \left( \nabla \cdot \mathbf{F}(\mathbf{r},t)\right) = 0$.
\end{enumerate}
Under conditions (i)--(iv), the solution to Equation \ref{eq:localbalance} becomes trivial:
\begin{equation}
\Delta T_{\text{eq}}(\mathbf{r},t) = -\frac{P_R(\mathbf{r})}{\lambda(\mathbf{r})} \overline{R}(t)
\label{eq:eqsolution}
\end{equation}
Where $\Delta T_{\text{eq}}(\mathbf{r},t)$ represents the equilibrium solution. The regional pattern of warming (i.e. regional warming normalized by the global average) is:
\begin{equation}
\frac{\Delta T_{\text{eq}}(\mathbf{r})}{\overline{\Delta T_{\text{eq}}}} = \frac{P_R(\mathbf{r})/\lambda(\mathbf{r})}{\overline{P_R(\mathbf{r})/\lambda(\mathbf{r})}}
\label{eq:equilibriumpattern}
\end{equation}

Equation \ref{eq:equilibriumpattern} has two interesting implications. First, under the four restrictive assumptions encoded in (i)---(iv), it implies that regional temperature anomalies scale exactly linearly with global temperature changes, where the linear coefficients are given by Equation \ref{eq:equilibriumpattern}. In other words, the ratio $\Delta T(\mathbf{r},t)/\overline{\Delta T}(t)$ is constant and only depends on the spatial distribution of the local feedback parameters and the forcing pattern, and is independent of the magnitude of the average forcing. Second, this analysis provides a clean framework to understand the specific factors and mechanisms that can make $\Delta T(\mathbf{r},t)/\overline{\Delta T}(t)$ variable in time and/or across scenarios. In the next subsections, we relax individual conditions in different combinations to isolate the specific mechanisms that can alter the warming pattern invariance. Table \ref{tab:roadmap} presents nine idealized cases that we consider in this work, where the above assumptions are relaxed either individually or in combination with others. The result in Equation \ref{eq:equilibriumpattern}, where all the four assumptions are retained, is labeled as \emph{Reference}.
\begin{table*}
\renewcommand{\arraystretch}{1.4} 
\centering
\caption{Summary of all the idealized experiments that we present in this work.}
\small
\begin{tabular}{lccccccc}
\hline
Case Label & Section & (i) Regime & (ii) Feedback & (iii) Forcing Pattern & (iv) Dynamics & Time Forcing & $\Delta T(\mathbf{r},t)/\overline{\Delta T}(t)$\\
\hline
\emph{Reference} & \ref{sec:idealized}\ref{subsec:localbalance} & Equilibrium & Linear & Invariant & --- & N/A & Constant \\
\emph{Exponential} & \ref{sec:idealized}\ref{subsec:transient} & Transient & Linear & Invariant & --- & Exponential & Constant \\
\emph{Abrupt} & \ref{sec:idealized}\ref{subsec:transient} & Transient & Linear & Invariant & --- & Constant & Variable \\
\emph{Overshoot} & \ref{sec:idealized}\ref{subsec:transient} & Transient & Linear & Invariant & --- & Gaussian & Variable \\
\emph{AlbedoFeedback} & \ref{sec:idealized}\ref{subsec:nonlinearfeedbacks} & Transient & Nonlinear & Invariant & --- & Exponential & Variable \\
\emph{AerosolForcing} & \ref{sec:idealized}\ref{subsec:forcingpattern} & Transient & Linear & Variable & --- & Exponential & Variable \\
\emph{DiffusionEq} & \ref{sec:idealized}\ref{subsec:dynamics} & Equilibrium & Linear & Invariant & Diffusion & N/A & Constant \\
\emph{DiffusionExp} & \ref{sec:idealized}\ref{subsec:dynamics} & Transient & Linear & Invariant & Diffusion & Exponential & Constant \\
\emph{DiffusionAbrupt} & \ref{sec:idealized}\ref{subsec:dynamics} & Transient & Linear & Invariant & Diffusion & Constant & Variable \\
\hline
\end{tabular}
\label{tab:roadmap}
\end{table*}

\subsection{The Transient Solution (i)}

\label{subsec:transient}

The first natural extension to the equilibrium model is the transient problem, i.e. we now relax assumption (i) and look at the time evolution before reaching equilibrium ($H(\mathbf{r},t) \ne 0$ during the transient period). This is highly relevant for the present climate and end-of-century projections, where the system is out-of-equilibrium due to slow ocean processes that have time scales on the order of millennia to adjust to  climate forcing \citep{held_probing_2010}. In this section, we show that the $\Delta T(\mathbf{r},t)/\overline{\Delta T}(t)$ ratio is still time- and scenario-invariant under exponential forcing whereas it evolves in time for abrupt and peak-and-decline forcing. In Appendix B, we show that the pattern invariance also holds approximately for linear and polynomial forcing.

Similarly to previous work that used energy balance models to study climate sensitivity and feedbacks \citep{bates_climate_2012,armour_time-varying_2013,rose_ocean_2009,merlis_interacting_2014,sodergren_energy_2018}, we parameterize $H(\mathbf{r},t)$ as the product of an effective heat capacity of the column $C(\mathbf{r})$ and the time rate of change of surface air temperature $\partial \Delta T(\mathbf{r},t)/\partial t$. This implies that the vertically-integrated energy content of the column is proportional to the surface air temperature $T(\mathbf{r},t)$ with proportionality coefficient $C(\mathbf{r})$.
With this parameterization, the general local energy balance (Equation \ref{eq:localbalance}) reads:
\begin{equation}
C(\mathbf{r})\frac{\partial \Delta T(\mathbf{r},t)}{\partial t} = \Delta N(\mathbf{r},t) = P_R(\mathbf{r})\overline{R}(t) + \lambda(\mathbf{r})\Delta T(\mathbf{r},t)
\label{eq:transient}
\end{equation}
Or, equivalently:
\begin{equation}
\frac{\partial \Delta T(\mathbf{r},t)}{\partial t} = -\frac{1}{\tau(\mathbf{r})}\Delta T(\mathbf{r},t) + f_R(\mathbf{r}) \overline{R}(t)
\label{eq:transientTimeScale}
\end{equation}
With $\tau(\mathbf{r})=-C(\mathbf{r})/\lambda(\mathbf{r})$ is the characteristic time scale and $f_R(\mathbf{r})=P_R(\mathbf{r})/C(\mathbf{r})$. The general solution to Equation \ref{eq:transientTimeScale} is:
\begin{equation}
    \Delta T( \mathbf{r},t ) = f_R(\mathbf{r})\int_{t_b}^t \exp{\left(\frac{t' - t}{\tau(\mathbf{r})}\right)}\overline{R}(t') \, dt'
    \label{eq:solutiontransient}
\end{equation}
Where $t'$ is introduced as a dummy variable for integration from the initial time $t_b$ to time $t$. 

It is instructive to further simplify Equation \ref{eq:solutiontransient} by studying specific forcing functions. We will consider three different cases: exponential forcing (\emph{Exponential}), constant forcing (\emph{Abrupt}) and an overshoot scenario, where forcing peaks and decline (\emph{Overshoot}). These three cases broadly mimic ssp585, abrupt4xCO2 and ssp119 presented in Section \ref{sec:CMIP6Evidence}, respectively. Two additional cases (linear and polynomial forcing) are presented in Appendix B. In the \emph{Exponential} case, we set $\overline{R}(t) = R_0\exp{\left(t/\tau_0\right)}$ and we integrate from $t_b = -\infty$ (when the forcing is zero and the system is in equilibrium with the forcing) to time $t$:

\begin{equation}
    \Delta T( \mathbf{r},t ) = \frac{R_0 f_R(\mathbf{r})}{1/\tau_0 + 1/\tau(\mathbf{r})}\exp{\left(\frac{t}{\tau_0}\right)}
    \label{eq:expsolution}
\end{equation}
With exponential forcing, temperatures at all locations grow exponentially at the same rate set by the time constant in the forcing function $\tau_0$. The scaling factor of each exponential depends on $\mathbf{r}$, and is determined by a combination of $R_0$, the heat capacity $C(\mathbf{r})$, the feedback $\lambda(\mathbf{r})$ and the forcing pattern $P_R({\mathbf{r}})$. This implies that $\Delta T(\mathbf{r},t)/\overline{\Delta T}(t) \ne 1$ (because of the different scaling factors), but is exactly independent of time and the forcing magnitude:
\begin{equation}
\frac{\Delta T(\mathbf{r},t)}{\overline{\Delta T}(t)} = \frac{A(\mathbf{r})}{\overline{A}}
\label{eq:transientpatternEXP}
\end{equation}
Where $A(\mathbf{r}) = f_R(\mathbf{r})/(1/\tau_0 + 1/\tau(\mathbf{r}))$ and $\overline{A}$ is the spatial average of $A(\mathbf{r})$. Note that Equation \ref{eq:transientpatternEXP} is of the same form as Equation \ref{eq:equilibriumpattern}, since neither $A(\mathbf{r})$ nor $\overline{A}$ depend on time $t$ and $A(\mathbf{r})/\overline{A}$ does not depend on the magnitude of the globally-averaged forcing. This implies that $\Delta T(\mathbf{r},t)/\overline{\Delta T}(t)$ is also time-invariant in the transient solution, but the values of the regional temperature pattern $\Delta T(\mathbf{r},t)/\overline{\Delta T}(t)$ are different from Equation \ref{eq:eqsolution} and  depend on $C(\mathbf{r}$) and $\tau_0$ in addition to $\lambda(\mathbf{r})$ and $P_R({\mathbf{r}})$. 

A second particular solution that we analyze is the constant-in-time forcing case $\overline{R}(t) = \overline{R}$, which is turned on at $t_b=0$ (\emph{Abrupt}). The general solution of Equation \ref{eq:solutiontransient} simplifies to:
\begin{equation}
\Delta T( \mathbf{r},t ) = \overline{R}f_R(\mathbf{r})\tau(\mathbf{r})\left[1-\exp\left(-\frac{t}{\tau(\mathbf{r})}\right)\right]
\label{eq:constantsolution}
\end{equation}
The constant forcing solution has drastically different features than the exponential forcing solution (Equation \ref{eq:expsolution}). Temperatures at different locations relax exponentially towards the equilibrium $-\overline{R}P_R(\mathbf{r})/\lambda(\mathbf{r})$ with \emph{different} time constants $\tau(\mathbf{r})$. This implies that the global average is a sum of exponentials and that $\Delta T(\mathbf{r},t)/\overline{\Delta T}(t)$  is dependent on time, but not on the forcing magnitude $\overline{R}$. 

The last forcing function that we consider is a Gaussian function $\overline{R}(t)=R_P\exp{(-(t - t_P)^2/(2 \sigma^2))}$ for \emph{Overshoot}. An analytical solution to this problem exists, although we omit the full expression because of the complexity arising from the square in the exponential. It has similar characteristics to the constant case solution, where each region has some exponential dependence with the $\tau(\mathbf{r})$ time scale, making the pattern variable in time during the transient adjustment.

We further illustrate the differences between \emph{Exponential}, \emph{Abrupt} and \emph{Overshoot} with numerical evaluations of Equation \ref{eq:solutiontransient}.  We follow the parsimonious demonstration by \cite{armour_time-varying_2013} and consider an idealized model where the planet is divided in three equal-area regions. The $\lambda(\mathbf{r})$ and $C(\mathbf{r})$ parameters are set to broadly mimic the warming patterns of land areas (\emph{Land}), the low-latitude oceans (\emph{Low}) and high-latitude oceans (\emph{High}), and their numerical values are reported in Table \ref{tab:threeregionsparams}. Specifically, the largest heat capacity and least stabilizing feedback parameter is found in the high-latitude oceans, whereas the lowest heat capacity is over land. The equilibrium climate sensitivity (ECS) of this idealized model is 3.6K, assuming the CO$_2$ doubling radiative forcing to be $\overline{R}_{2\times} = 3.7$ W m$^{-2}$, uniformly distributed across the three regions ($P_R(\mathbf{r}) = 1$).
\begin{table}[bthp]
  \footnotesize
  \def\arraystretch{1.25}%
  \centering
  \caption{Model parameters for the transient three-region model, as in \cite{armour_time-varying_2013}. The heat capacity is expressed in terms of effective water depth $h$, i.e. $C(\mathbf{r}) = \rho_w c_w h(\mathbf{r})$, where $\rho_w$ and $c_w$ are the density and specific heat capacity of water, respectively.}
    \begin{tabular}{lccccc}
       \hline
    Parameter & Symbol & \emph{Land} & \emph{Low} & \emph{High} \\
       \hline
    Effective water depth (m) & $h(\mathbf{r})$ & 10 & 150 & 1500 \\
    Local Feedback (W m$^{-2}$ K$^{-1}$) & $\lambda(\mathbf{r})$ & -0.86 & -2.0 &  -0.67 \\  
    Area Fraction (-) & $w(\mathbf{r})$ & 1/3 & 1/3 & 1/3 \\
    Forcing Pattern (-) & $P_R(\mathbf{r})$ & 1 & 1 & 1 \\
    Time constant (years) & $\tau(\mathbf{r})$ & 1.5 & 9.5 & 280 \\
    \hline
    \end{tabular}%
    \label{tab:threeregionsparams}
\end{table}%

\begin{figure*}[t]
\centerline{\includegraphics[width=0.8\textwidth]{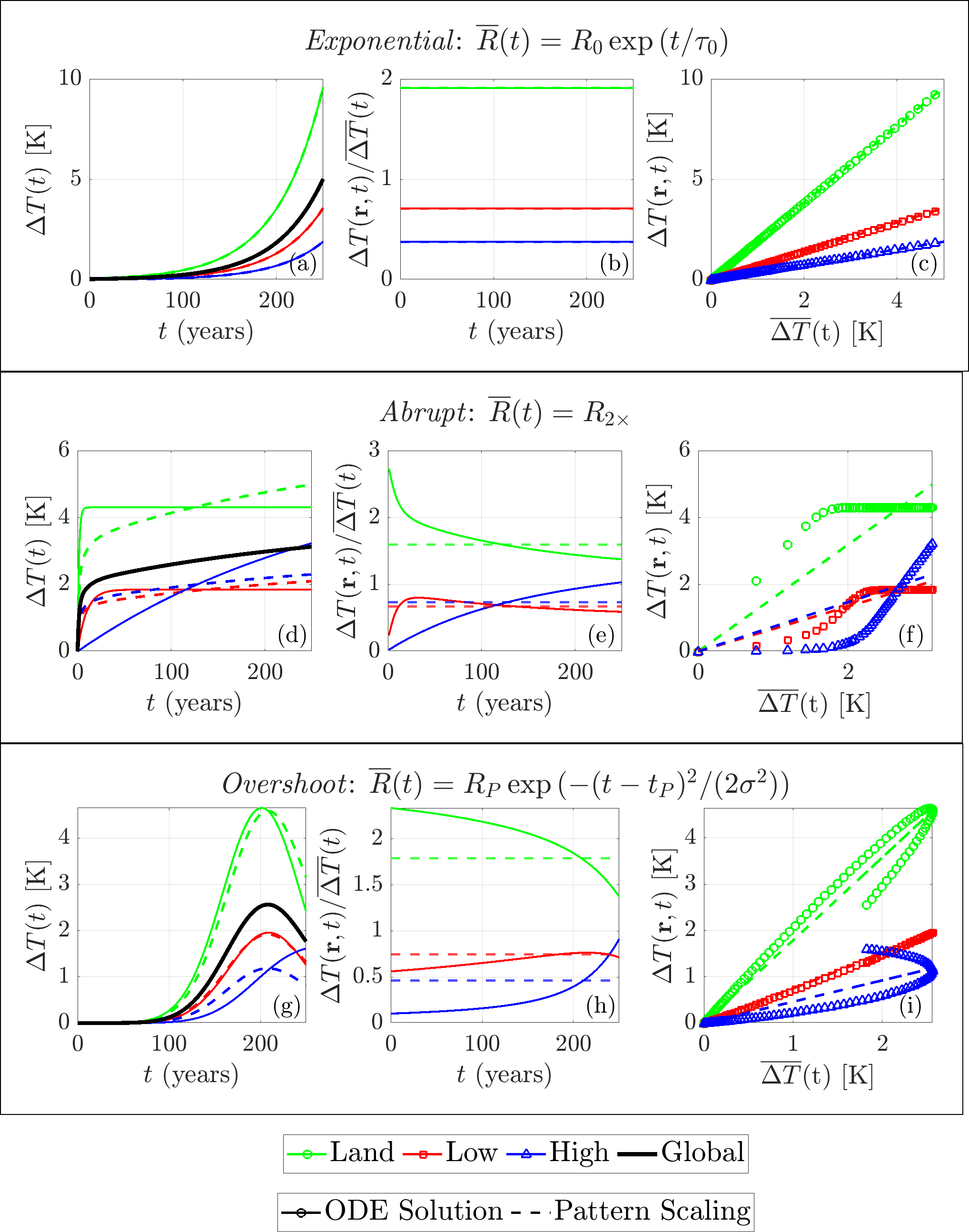}}
  \caption{Demonstration of evolving warming patterns ($\Delta T(\mathbf{r},t)/\overline{\Delta T}(t)$) under different forcings, with constant feedback parameters, no changes in dynamic heating rates and homogeneous forcing in space: \emph{Exponential} (top), \emph{Abrupt} (middle) and \emph{Overshoot} (bottom). The left column shows the comparison between the local energy balance (Equation \ref{eq:transient}) ODE solution (solid) and temperatures scaled with the best-fit $\Delta T(\mathbf{r},t)/\overline{\Delta T}(t)$ (dashed); the middle column  shows the time evolution of $\Delta T(\mathbf{r},t)/\overline{\Delta T}(t)$ (solid) and the best-fit time-invariant estimate from linear regression (dashed); the right column shows the $\Delta T(\mathbf{r},t)$ on $\overline{\Delta T}(t)$ linear regression (markers from the ODE solution, dashed line is the best-fit line with zero intercept).}
  \label{fig:transientevaluations}
\end{figure*}

Figure \ref{fig:transientevaluations} shows 250-year integrations of Equation \ref{eq:transient} with the parameters of Table \ref{tab:threeregionsparams} for \emph{Exponential}, \emph{Abrupt} and \emph{Overshoot}. The time horizon that we simulate is similar to typical integrations from CMIP6 over the historical period and future end-of-century projections (1850--2100). The left column shows the temperature evolution over the 250 years of integration in the three regions along with the global average. Dashed lines in the left column represent the regional warming calculated by scaling the global average (thick line) with a time-invariant $\Delta T(\mathbf{r},t)/\overline{\Delta T}(t)$ computed by regressing $\Delta T(\mathbf{r},t)$ on $\overline{\Delta T}(t)$. This is the typical pattern scaling calculation done in climate impact assessments \citep{tebaldi_pattern_2014}, that perfectly reproduces the climate model output only if $\Delta T(\mathbf{r},t)/\overline{\Delta T}(t)$ is exactly time invariant. The middle column shows the time evolution of $\Delta T(\mathbf{r},t)/\overline{\Delta T}(t)$ along with the best-fit values obtained from the $\Delta T(\mathbf{r},t)$ over $\overline{\Delta T}(t)$ linear regression. The right column shows such regression along with the best fit line (dashed). The slopes of the dashed lines in the right column are the constant values represented with dashed lines in the middle column, which are used to scale the global average in the left column.

For \emph{Exponential} (top row in Figure \ref{fig:transientevaluations}), we select $R_0=0.0573$ W m$^{-2}$ and $\tau_0 = 50$ years to have $\overline{R}(t=250)$ = 8.5 W m$^{-2}$, similarly to ssp585. The radiative forcing is uniform across regions ($P_R(\mathbf{r}) = 1$).
Under exponential forcing, the pattern scaling calculation is indistinguishable from the ODE output (Figure \ref{fig:transientevaluations}a) as the pattern is exactly time-invariant (Figure \ref{fig:transientevaluations}b). The local-to-global regression is perfectly linear.

For \emph{Abrupt} (middle row in Figure \ref{fig:transientevaluations}), we set a spatially homogeneous $\overline{R}(t)=\overline{R}_{2\times}=3.7$ W m$^{-2}$, which is roughly equivalent to doubling the preindustrial CO$_{2}$ concentration in the atmosphere. The response of the three regions to  constant forcing shows the emergence of the different time scales in the system. The \emph{Land} region reaches equilibrium almost immediately ($<5$ years), whereas the \emph{High} region is still far from equilibrium even at the end of the integration time (250 years). The global average reflects both these behaviors: the planet warms very quickly right after the forcing is turned on (driven by the \emph{Land} region), and keeps warming at a slower pace towards the end of the simulation (driven by the \emph{High} region). This is a very similar behavior to what is observed in CMIP6 experiments where concentrations of CO$_2$ are abruptly doubled or quadrupled \citep{held_probing_2010}. Pattern scaling is not a good approximation of the system, as we derived analytically in Equation \ref{eq:constantsolution}, because the global average grows as a sum of exponentials with different time scales and thus does not reflect the time evolution of any individual region. This is also evident from panel (e) in Figure \ref{fig:transientevaluations}, which clearly shows that $\Delta T(\mathbf{r},t)/\overline{\Delta T}(t)$ is far from being time-invariant in \emph{Abrupt}, and panel (f), which shows that $\Delta T(\mathbf{r},t)$ plotted against $\overline{\Delta T}(t)$ does not align on a straight line.

For \emph{Overshoot}, we set $R_P=4$ W m$^{-2}$ (peak forcing), $t_P=200$ years (timing of the peak) and $\sigma=42$ years (set to have forcing at year 250 to be half of the peak forcing).  As in the constant forcing case, the warming pattern evolves in time: the \emph{Land} region peaks and declines faster than the global average, because the global average is influenced by the temperature in the \emph{High} region that increases even after the forcing declines at $t_P=200$ years. As a result, pattern scaling estimates are not a good approximation of the real system for the \emph{High} region, and are also slightly incorrect in the two other regions because of the influence from the global average. For instance, consider a global temperature anomaly threshold of $\overline{\Delta T} = 2$K, which is crossed twice (before and after the peak). At the time of crossing, the \emph{High} region shows two separate and very different temperature anomalies (0.5K and 1.5K), which correspond to two very different patterns (0.25 and 0.75, respectively). This result highlights the problem of using one single time-invariant pattern in \emph{Overshoot}, where the regional temperature obtained by scaling the global average with the best-fit pattern for the \emph{High} region ($\Delta T = 0.8$K) is not capturing the correct values either before (0.5K) or after the peak (1.5K).

To summarize, two conditions are sufficient to make the $\Delta T(\mathbf{r},t)/\overline{\Delta T}(t)$ ratio time-dependent during the transient adjustment, if we still retain simplifications (ii)-(iv): (1) different $\lambda/C$ in different regions \emph{and} (2) a time forcing that is not exponential. This is because linear ODE systems have exponential eigenfunctions in time (i.e., a function that when applied as a forcing creates a response of the same shape)  as noted by \cite{raupach_exponential_2013} in the context of the carbon-climate system. For linear and polynomial forcing functions, the warming pattern is also invariant after an initial transient where the system adjusts and locks in with the forcing (Appendix B), and pattern scaling is still a good approximation of the real system.

\subsection{Nonlinear Feedbacks (ii)}

\label{subsec:nonlinearfeedbacks}

In this section we illustrate an additional mechanism that breaks the time- and scenario-invariance of $\Delta T(\mathbf{r},t)/\overline{\Delta T}(t)$, related to the nonlinear surface albedo feedback in the Arctic region. We isolate the surface albedo feedback for the sake of illustration; similar arguments could be made for other nonlinear feedbacks.

In Section \ref{sec:idealized}\ref{subsec:transient}, we assumed that the feedback parameter $\lambda(\mathbf{r})$ lumped the effects of both shortwave and longwave feedbacks, and that it was time- and forcing-invariant. To study the surface albedo feedback, it is useful to separate the shortwave and longwave contributions:
\begin{equation}
\Delta N = \Delta \text{ASR} - \Delta \text{OLR}
\end{equation}
Where ASR and OLR are the shortwave (Absorbed Solar Radiation) and longwave (Outgoing Longwave Radiation) components of the total energy balance. The connection with the linearization in assumption (ii) is the following:
\begin{equation}
    \Delta N = \underbrace{\lambda_S \Delta T}_{\Delta\text{ASR}} \quad - \quad \underbrace{(-R - \lambda_L \Delta T)}_{\Delta\text{OLR}} = R + \lambda \Delta T
    \label{eq:DNdecomposition}
\end{equation}
Where $\lambda=\lambda_S+\lambda_L$, $\lambda_S$ and $\lambda_L$ are the shortwave and longwave feedbacks, respectively, and we assumed that the radiative forcing operates in the longwave (e.g. CO$_2$), without loss of generality. We further decompose the shortwave feedback into the surface albedo feedback ($\lambda_{S,A}$) and shortwave cloud feedback ($\lambda_{S,C}$), i.e. $\lambda_S=\lambda_{S,A}+\lambda_{S,C}$. We derive what $\lambda_{S,A}$ ought to be by studying changes in ASR as a function of changing surface albedo.

The change in ASR from a reference climate with surface albedo $\alpha_{PI}$ and a perturbed climate with surface albedo $\alpha$ is:
\begin{equation}
\Delta \text{ASR} = S(1-\alpha) - S(1-\alpha_{PI}) = -S\Delta\alpha
\end{equation}
Where $S$ is the incoming solar radiation at the surface. If $\alpha$ is a linear function of surface temperature, i.e. $\Delta\alpha = k\Delta T$ (where $k$ is negative for melting sea ice), the net change in ASR due to changing surface albedo is $\Delta \text{ASR}=-kS\Delta T$. Combining with Equation \ref{eq:DNdecomposition}, this implies that $\lambda_S=-kS$ is constant and positive. In the Arctic, the assumption that $\alpha$ is linearly related with temperature is expected to be violated, because $\alpha$ is bounded between the sea-ice albedo ($\alpha_i$) and the ocean albedo ($\alpha_o$) if the Arctic becomes sea-ice free. Early energy balance models assumed the surface albedo temperature dependence to be a step-function \citep{held_simple_1974, north_analytical_1975}, a piecewise linear function \citep{held_albedo_1981} or a hyperbolic tangent function \citep{merlis_interacting_2014}. All these three functional forms are bounded between $\alpha_i$ and $\alpha_o$, and they differ in how the transition between the two occurs. In this work, we follow \citet{merlis_interacting_2014} and specify an hyperbolic tangent function for the surface albedo temperature dependence:
\begin{equation}
\alpha(\Delta T) = \frac{\alpha_o + \alpha_i}{2} + \frac{\alpha_o - \alpha_i}{2}\tanh{\left(\frac{\Delta T-\Delta T_0}{h_T}\right)}
\label{eq:albedo}
\end{equation}
Where $\Delta T_0=10$ K is the temperature increase needed to have $\alpha(T)$ equal to the average of $\alpha_o$ and $\alpha_i$, and $h_T=6$ K sets how rapidly the transition from $\alpha_i$ to $\alpha_o$ occurs (we use the same value as in \cite{merlis_interacting_2014}). Note that our goal is to demonstrate this mechanism qualitatively, so that the specific choice for the parameters of Equation \ref{eq:albedo} is realistic but not necessarily tuned to observations and/or climate models.

Consider now a simple model where Earth is divided in two regions, named \emph{Arctic} and \emph{ROW} (Rest Of the World). If this model is subject to the same exponential forcing as in Figure \ref{fig:transientevaluations} with assumptions (ii)--(iv), $\Delta T(\mathbf{r},t)/\overline{\Delta T}(t)$ is time- and scenario-invariant as we have demonstrated in the previous Section. We break assumption (ii) by setting the feedback parameter in the \emph{Arctic} region to be the sum of a fixed feedback (the sum of $\lambda_{S,C}$ and $\lambda_L$) and the surface albedo feedback that varies with temperature (i.e., $\lambda_{S,A} = \lambda_{S,A}(\Delta T)$). The \emph{ROW} has a fixed feedback parameter that accounts for both shortwave and longwave effects. We set the fixed feedback to -1.25 W m$^{-2}$ K$^{-1}$ for both \emph{Arctic} and \emph{ROW}. The surface albedo feedback is $-S\Delta\alpha$, where $\alpha$ is calculated interactively with Equation \ref{eq:albedo} and S is set to 100 W m$^{-2}$ (roughly consistent with CMIP6 data for the annually-averaged surface downward radiation in the Arctic). We solve the energy balance equation numerically given the nonlinearity introduced by the surface albedo feedback, with heat capacity corresponding to 100m of water and the same exponential forcing as in the previous Section. For the global average calculations, we assign area weights of 0.02 and 0.98 to \emph{Arctic} and \emph{ROW}, respectively, corresponding to the surface area poleward of 70$^\circ$N and the rest of the planet.

\begin{figure*}[t]
\centerline{\includegraphics[width=\textwidth]{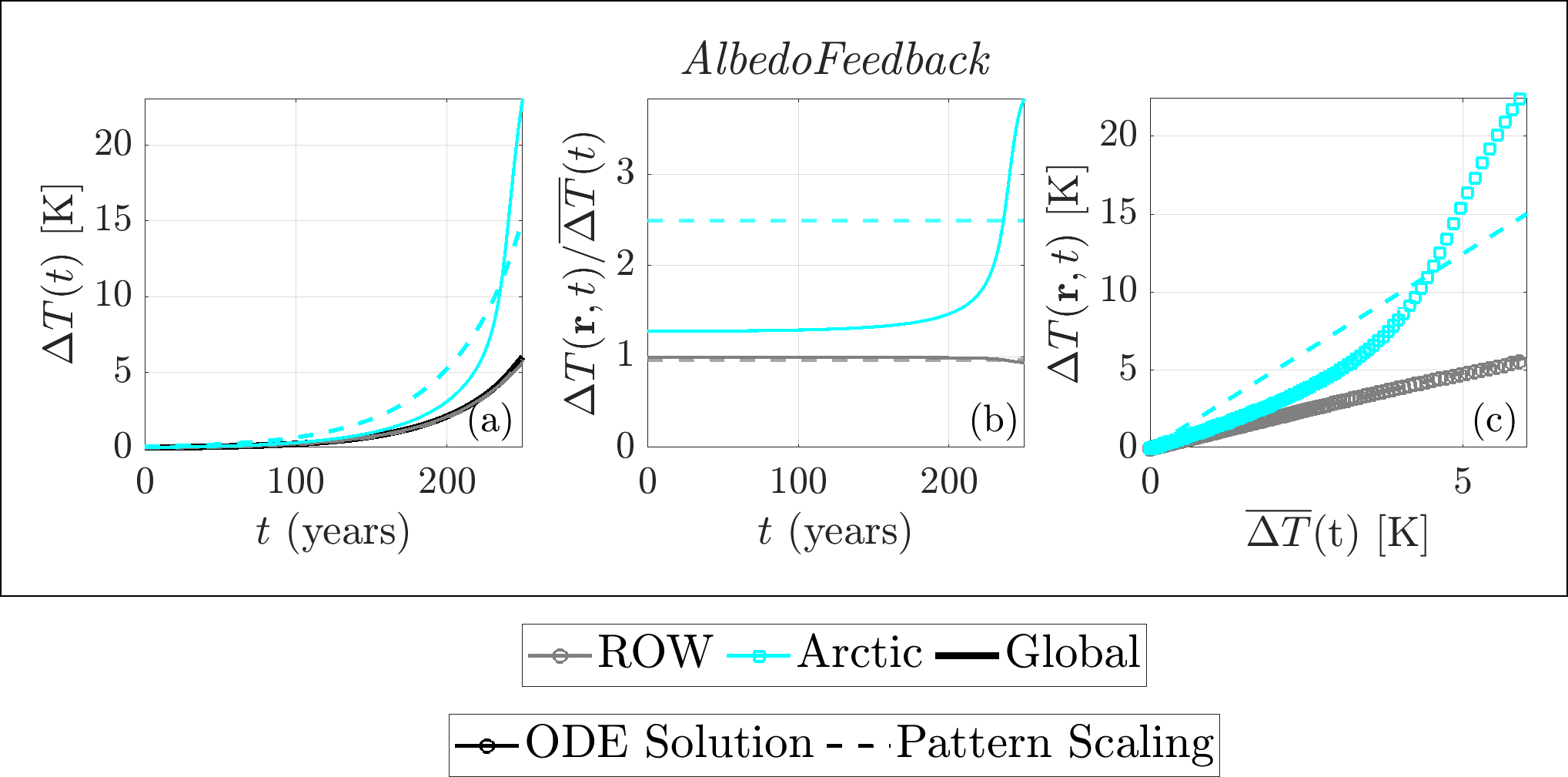}}
  \caption{Demonstration of the effect of nonlinear feedbacks on the warming pattern in the \emph{AlbedoFeedback} idealized case (no changes in heat transport, homogeneous exponential forcing). The specific forcing parameters are described in the main text. The left column shows the comparison between the ODE solution (solid) and temperatures scaled with the best-fit $\Delta T(\mathbf{r},t)/\overline{\Delta T}(t)$ (dashed); the middle column  shows the time evolution of $\Delta T(\mathbf{r},t)/\overline{\Delta T}(t)$ (solid) and the best-fit time-invariant estimate from linear regression (dashed); the right column shows the $\Delta T(\mathbf{r},t)$ on $\overline{\Delta T}(t)$ linear regression (markers from the ODE solution, dashed line is the best-fit line with zero intercept).}
  \label{fig:nonlinearfeedbacks}
\end{figure*}

Figure \ref{fig:nonlinearfeedbacks} shows
the 250-year integrations for this simple case (\emph{AlbedoFeedback}). The \emph{Arctic} warming pattern remains almost constant for most of the simulation but grows considerably towards the end when the nonlinear part of the albedo function is triggered. 
This gives a distinctive concavity to the global-to-local regression in Panel (c), which is related to the acceleration of sea ice melting (or albedo decrease) as temperature grows. For even higher temperatures, not reached in this simulation, we expect a saturation of this mechanism because $\alpha$ is bounded to $\alpha_o$, and a change in concavity in the global-to-local regression. Note that scaling the global temperature with the regressed linear pattern fails to capture both the early and late temperature increase in Figure \ref{fig:nonlinearfeedbacks}a, even before the nonlinear part of the albedo function is triggered. The largest pattern scaling error in \emph{AlbedoFeedback} is more than 5 K at the end of the simulation, which is entirely related to the nonlinear feedback effect. 

\subsection{Non-constant Forcing Patterns (iii)}

\label{subsec:forcingpattern}

We now consider a simple problem where the forcing pattern varies in time, i.e. $P_R = P_R(\mathbf{r},t)$, relaxing assumption (iii). The problem is set up to mimic the cooling effect of tropospheric aerosols. The starting point is the same \emph{Exponential} case that we discussed in Section \ref{sec:idealized}\ref{subsec:transient} (transient, linear feedbacks, no changes in dynamic heating rates). We introduce spatial inhomogeneity through the time scale that regulates the forcing, which we allow to vary in space, i.e. $\tau_0=\tau_0(\mathbf{r})$. The globally-averaged radiative forcing (as well as the globally-averaged temperature response) becomes a sum of exponentials that accounts for all the time scales in the different region-dependent forcings, whereas each individual region responds to the single region-specific time scale. Note that this is conceptually similar to the \emph{Abrupt} case, where the time-dependence of $\Delta T(\mathbf{r},t)/\overline{\Delta T}(t)$ was introduced via $\lambda/C$ rather than through the forcing pattern. 

To demonstrate this mechanism numerically, we set up a simple model (\emph{AerosolForcing}) with three regions, labeled \emph{LateEmitter}, \emph{EarlyEmitter} and \emph{ROW}. We impose an exponential ($R(t,\mathbf{r}) = R_0(\mathbf{r})\exp{\left(t/\tau_0(\mathbf{r})\right)}$) negative radiative forcing over \emph{LateEmitter} and \emph{EarlyEmitter}, mimicking the cooling effect of tropospheric aerosols (e.g., sulfate). Specifically, the exponential forcing parameter $R_0$ is set to reach -2.0 W m$^{-2}$ at year 250 in both regions, with different exponential rates for \emph{EarlyEmitter} and \emph{LateEmitter} ($\tau_0=25$ years and $\tau_0=75$ years, respectively). We also assume zero forcing over \emph{ROW}. The two time scales represent two different regions emitting aerosols with different pathways.  We set the feedback parameters and the heat capacities to the same values for all regions  ($h=100$ m and $\lambda = -0.86$ W m$^{-2}$), and we assume that \emph{EarlyEmitter} and \emph{LateEmitter} cover 10\% of Earth's area each for the global average calculation.

\begin{figure*}[t]
\centerline{\includegraphics[width=\textwidth]{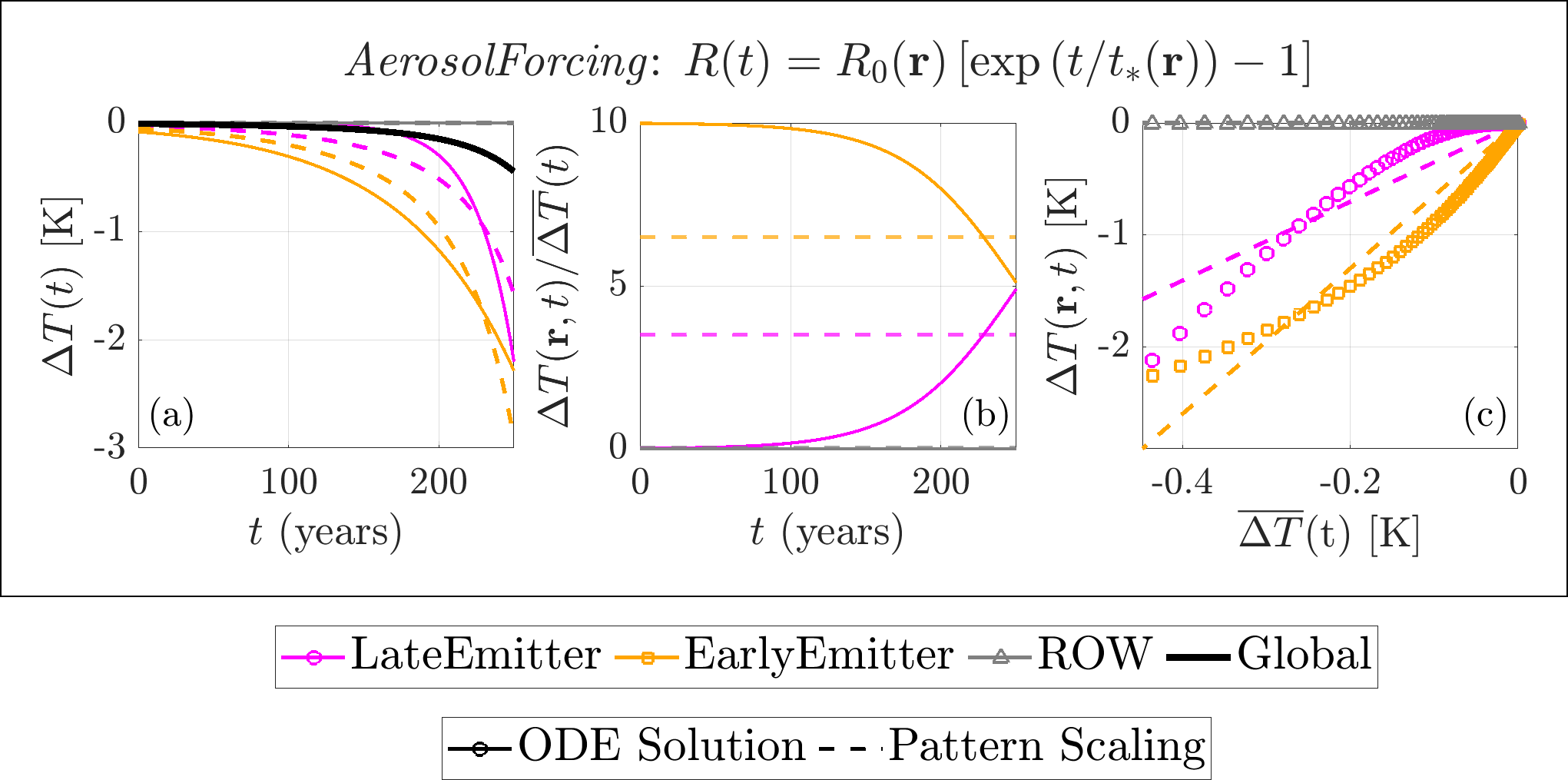}}
  \caption{Demonstration of the effect of a time-varying forcing pattern on the warming pattern in the \emph{AerosolForcing} idealized case (no changes in heat transport, homogeneous exponential forcing with region-dependent $\tau_0$). The specific forcing parameters are described in the main text. The left column shows the comparison between the ODE solution (solid) and temperatures scaled with the best-fit $\Delta T(\mathbf{r},t)/\overline{\Delta T}(t)$ (dashed); the middle column  shows the time evolution of $\Delta T(\mathbf{r},t)/\overline{\Delta T}(t)$ (solid) and the best-fit time-invariant estimate from linear regression (dashed); the right column shows the $\Delta T(\mathbf{r},t)$ on $\overline{\Delta T}(t)$ linear regression (markers from the ODE solution, dashed line is the best-fit line with zero intercept).}
  \label{fig:inhomogeneous}
\end{figure*}

Figure \ref{fig:inhomogeneous} shows the numerical evaluation of \emph{AerosolForcing} in a similar format to Figure \ref{fig:transientevaluations}.
As expected, scaling the global average with the best-fit time-invariant pattern provides inaccurate estimates of the true regional temperatures evolution from the ODEs, somewhat similarly to the constant forcing case with homogeneous forcing described in the previous Section. We expect this to be potentially significant for the historical and projected aerosol emissions in different regions of the world, that had (and will likely have) the timing of their peak aerosol forcing shifted by a few decades \citep{bauer_turning_2022}. Note that the forcing pattern has a large variation in time (from 10 to 5 in 250 years), although the relative mismatch between pattern scaling calculations and the true ODE solution is comparatively small. This is because the global temperature decrease (denominator in the pattern ratio) is less than 0.5K during the simulation, which makes the relative patterns comparatively large.

\subsection{Changes in Dynamic Heating Rates (iv)}

\label{subsec:dynamics}

In this section we study the effect of relaxing assumption (iv), i.e. $\Delta \left( \nabla \cdot \mathbf{F}(\mathbf{r},t)\right) \ne 0$. Changes in $\nabla \cdot \mathbf{F}$ introduce substantial complexity to the system, but as long as those changes can be represented as a time-independent linear operator the problem remains amenable to a rather simple discussion. For simplicity, we consider $T(\mathbf{r}) = T(\phi)$ representing the mean temperature at latitude $\phi$, i.e. we average out the longitudinal dependence and we study the changes in meridional heat transport (MHT) across different latitudes. We parameterize $\Delta \left( \nabla \cdot \mathbf{F}(\mathbf{r},t)\right)$ with \cite{sellers_global_1969} diffusion, which represent changes in dynamical heating rates as a diffusive process directed down the gradient of temperature anomaly:
\begin{equation}
\Delta \left( \nabla \cdot \mathbf{F}(\phi,t)\right) = \frac{D}{\cos{\phi}}\frac{\partial}{\partial \phi}\left(\cos{\phi}\frac{\partial \Delta T(\phi)}{\partial \phi}\right)
\label{eq:dynchanges}
\end{equation}
Where $D$ is the climate diffusivity parameter in m$^2$ s$^{-1}$, assumed to be temperature-independent as in the original \cite{sellers_global_1969} work and subsequent diffusion models \citep{rose_ocean_2009,merlis_interacting_2014}. 
If we relax assumption (iv) while retaining (i)-(iii), and parameterize $\Delta \left( \nabla \cdot \mathbf{F}(\phi,t)\right)$  with Sellers diffusion, the local energy balance equation becomes:
\begin{equation}
 P_R(\phi)\overline{R}(t) + \lambda(\phi)\Delta T(\phi,t) + \frac{D}{\cos{\phi}}\frac{\partial}{\partial \phi}\left(\cos{\phi}\frac{\partial \Delta T(\phi)}{\partial \phi}\right) = 0
\label{eq:ebdynchanges}
\end{equation}

Equation \ref{eq:ebdynchanges} is a linear boundary value problem with no-flux boundary conditions at $\phi=\pm 90^{\circ}$. This implies that the equilibrium regional temperature anomalies scale linearly with the forcing input, and so does the global average, making $\Delta T(\phi,t)/\overline{\Delta T}(t)$ constant for any forcing. This does not mean that dynamic heating rates are also time invariant, but rather that $\Delta \left( \nabla \cdot \mathbf{F}(\phi,t)\right)$ scales linearly with the forcing. In other words, the regional pattern $\Delta T(\phi,t)/\overline{\Delta T}(t)$ remains unique at equilibrium even if we have changes in dynamic heating. 

As a concrete example, consider the same model described in Table \ref{tab:threeregionsparams}, where we now assign latitude bands to the three regions (\emph{DiffusionEq}). Specifically, we assign the \emph{Land} parameters to the midlatitudes (18$^\circ$S-54$^\circ$S, 18$^\circ$N-54$^\circ$N), the \emph{Low} parameters to the tropical region (18$^\circ$S-18$^\circ$N) and the \emph{High} parameters to high-latitudes (54$^\circ$S-90$^\circ$S, 54$^\circ$N-90$^\circ$N). The specific numbers are set to have polar-amplified warming without necessarily capturing all the details of realistic regional warming. 

The equilibrium warming calculated numerically with a realistic value of $D=0.55$ W m$^{-2}$ K$^{-1}$ is reported in Table \ref{tab:dyneqsolution}, for two different forcing values that roughly correspond to 2$\times$CO$_2$ and 4$\times$CO$_2$ (3.7 W m$^{-2}$ and 7.4 W m$^{-2}$, respectively). Given the linear ODE structure, doubling the forcing doubles the temperature response and the dynamical changes, keeping the polar-amplified pattern constant. The equilibrium warming is also smoother than the one that would occur with no changes in heat transport ($-R/\lambda$). For instance, the high-latitude equilibrium warming with $D=0$ and $R=3.7$ W m$^{-2}$ is $-R/\lambda=5.5$ K, whereas with $D=0.55$ W m$^{-2}$ K$^{-1}$ is only 4.2 K (Table \ref{tab:dyneqsolution}). This is because the high latitudes warm more than the tropics, reducing the average equator-to-pole temperature gradient and the dynamic heating rates at the pole. An alternative and equivalent view of this behavior comes from Equation \ref{eq:ebdynchanges}, which is effectively a diffusive system for the scalar $\Delta T(\phi)$.

\begin{table*}
  \footnotesize
  \def\arraystretch{1.25}%
  \centering
  \caption{Equilibrium solution of the local energy balance model (Equation \ref{eq:ebdynchanges}) with constant $D=0.55$ W m$^{-2}$ K$^{-1}$, under 2$\times$CO$_2$ forcing ($R=3.7$ W m$^{-2}$) and 4$\times$CO$_2$ forcing ($R=7.4$ W m$^{-2}$)}
\begin{tabular}{ll|ccc|ccc}
\hline
\multicolumn{2}{c|}{} & \multicolumn{3}{c|}{2$\times$CO$_2$} & \multicolumn{3}{c}{4$\times$CO$_2$} \\
\hline
$\phi$ & $\lambda(\phi)$ & $\Delta T(\phi)$ & $\Delta \left( \nabla \cdot \mathbf{F}(\phi)\right)$ & $\Delta T(\phi)/\overline{\Delta T}$ & $\Delta T(\phi)$ & $\Delta \left( \nabla \cdot \mathbf{F}(\phi)\right)$ & $\Delta T(\phi)/\overline{\Delta T}$ \\
--- & W m$^{-2}$ K$^{-1}$ & K & W m$^{-2}$ & K/K & K & W m$^{-2}$ & K/K \\
\hline
54S--90S & -0.67 & 3.88 & -1.10 & 1.17 & 7.76 & -2.20 & 1.17 \\
18S--54S & -0.86 & 3.47 & -0.72 & 1.04 & 6.93 & -1.44 & 1.04 \\
18S--18N & -2.00 & 2.77 & 1.84 & 0.83 & 5.54 & 3.68 & 0.83 \\
18N--54N & -0.86 & 3.47 & -0.72 & 1.04 & 6.93 & -1.44 & 1.04 \\
54N--90N & -0.67 & 3.88 & -1.10 & 1.17 & 7.76 & -2.20 & 1.17 \\
\hline
\end{tabular}
\label{tab:dyneqsolution}
\end{table*}

\begin{figure*}[t]
\centerline{\includegraphics[width=\textwidth]{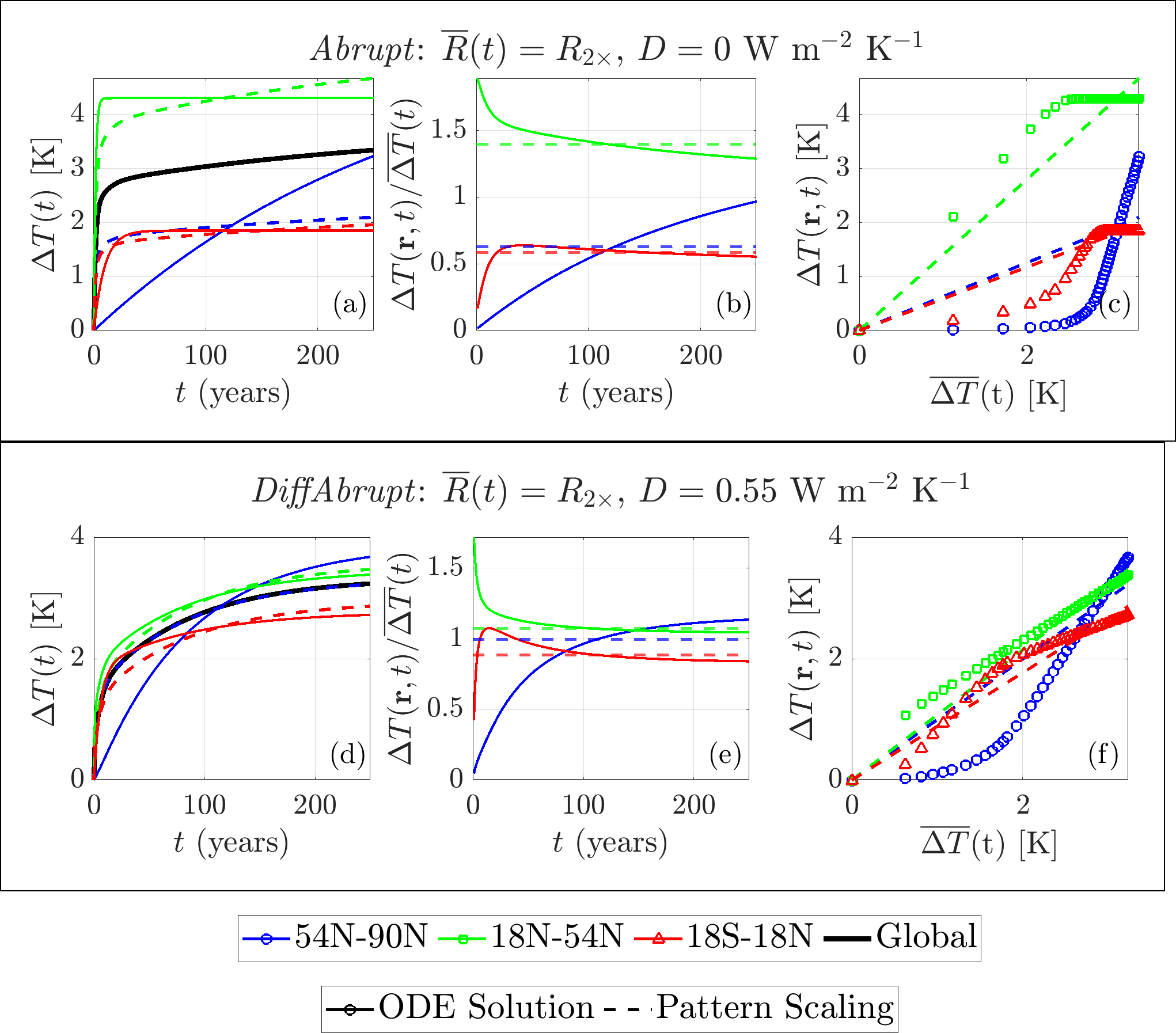}}
  \caption{Demonstration of the effect of diffusive dynamics on the warming pattern in the \emph{DiffAbrupt} idealized case (homogeneous constant forcing, linear feedbacks). The specific forcing parameters are described in the main text. The left column shows the comparison between the ODE solution (solid) and temperatures scaled with the best-fit $\Delta T(\mathbf{r},t)/\overline{\Delta T}(t)$ (dashed); the middle column  shows the time evolution of $\Delta T(\mathbf{r},t)/\overline{\Delta T}(t)$ (solid) and the best-fit time-invariant estimate from linear regression (dashed); the right column shows the $\Delta T(\mathbf{r},t)$ on $\overline{\Delta T}(t)$ linear regression (markers from the ODE solution, dashed line is the best-fit line with zero intercept).}
  \label{fig:dynchanges}
\end{figure*}

If we also relax assumption (i) and look at the transient adjustment ($\partial/\partial t \ne 0$), the behavior depends on the forcing function as discussed in the previous Sections. The exponential case is still pattern-invariant because the diffusion operator is linear, which means that an exponential function is still an eigenfunction of the system. The constant forcing case (\emph{DiffusionAbrupt}) is presented in Figure \ref{fig:dynchanges}, which shows the comparison between turning diffusion off ($D=0$) and using a typical value of the diffusivity coefficient $D=0.55$ W m$^{-2}$ K$^{-1}$, for the same problem described above with 5 latitude bands. The $D=0$ case is almost equivalent to \emph{Abrupt}, except for the area weighting coefficients that are not prescribed but set by the definition of the latitude bands. 
Figure \ref{fig:dynchanges} shows that MHT effectively smooths some of the regional differences related to different heat capacities and feedbacks in different regions. Temperature in the 18N--54N latitude band, for instance, keeps increasing throughout the entire integration despite the really short equilibration time scale, because neighboring regions with higher heat capacities exchange heat meridionally. Purely diffusive changes in MHT effectively alleviate the pattern scaling mismatch and make the $\Delta T(\mathbf{r},t)/\overline{\Delta T}(t)$ ratio closer to a constant.  In the limit of $D \to \infty$, temperature at all latitudes would collapse on a single curve (equivalent to the global average) and the regression on the rightmost panel would be perfectly linear.

\section{Interpretation of the idealized experiments in the context of CMIP6}

\label{sec:interpretation}

We seek to understand if the theoretical arguments that we discussed in Section \ref{sec:idealized} provide useful information for interpreting the CMIP6 results that we summarized in Section \ref{sec:CMIP6Evidence}. We proceed by analyzing assumptions (i)---(iii) and present additional data to study to what extent the three assumptions are met in CMIP6 experiments. For assumption (iv), we rely on qualitative arguments based on previous literature acknowledging that a comprehensive and more quantitative understanding is a promising avenue for future work.

\subsection{Transient and Equilibrium Regime (i)}

Experiments from CMIP6 coupled models are inherently transient, since ocean processes have long time scales to reach equilibrium. We analyze the same CMIP6 experiments that we have presented in Section \ref{sec:CMIP6Evidence}, which are qualitatively similar to \emph{Exponential} (ssp585), \emph{Overshoot} (ssp119) and \emph{Abrupt} (abrupt-4xCO2). Note that $\overline{R}(t)$ in historical and ssp585 is roughly exponential because CO$_2$ concentrations grow exponentially \citep{meinshausen_shared_2020}, and radiative forcing is described by the following expression (for the range of CO$_2$ concentrations in historical and ssp585):
\begin{equation}
    \overline{R}(t) = \left[d_1 + a_1(C-C_0)^2+b_1(C-C_0)\right]\log\left(\frac{C}{C_0}\right)
    \label{eq:lblrf}
\end{equation}
where $d_1$, $a_1$ and $b_1$ are constants, and $C_0$ and $C$ are the CO$_2$ concentrations during preindustrial and at time $t$, respectively. The specific values of $d_1$, $a_1$ and $b_1$ in Equation \ref{eq:lblrf} are fitted from line-by-line radiation calculations and are described in \citet{meinshausen_shared_2020}. The quadratic prefactor is responsible for retaining an approximately exponential $\overline{R}(t)$ from an exponential increase of CO$_2$ concentrations, despite the logarithmic term in Equation \ref{eq:lblrf}.

\begin{figure*}[t]
\centerline{\includegraphics[width=\textwidth]{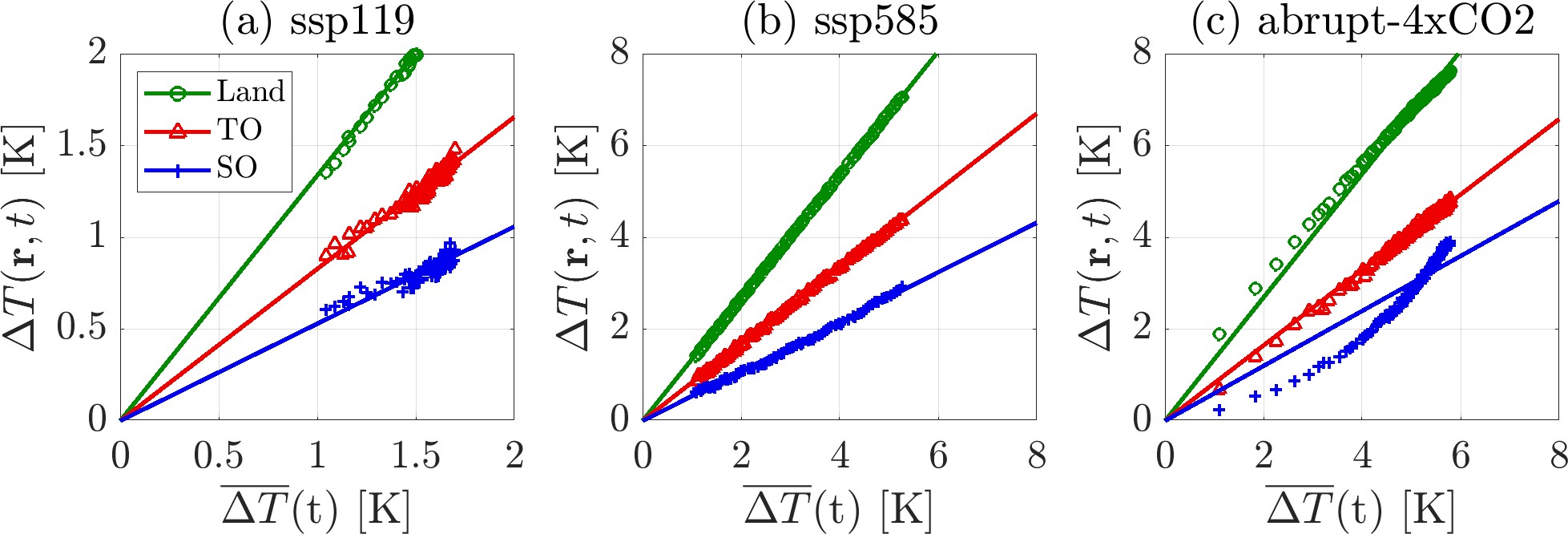}}
  \caption{Local-to-global regressions for Land, TO and SO in (a) ssp119, (b) ssp585 and (c) abrupt-4xCO2. Data are from the CMIP6 multimodel average. Note the different $x$ and $y$ axes scales in panel (a) compared to (b) and (c).}
  \label{fig:cmipLR}
\end{figure*}

Figure \ref{fig:cmipLR} shows that the global-to-regional relationships qualitatively agree with the theory presented in Section \ref{sec:idealized}. Specifically, ssp585 shows a linear relationship for the three regions considered (as in \emph{Exponential}), whereas some deviations from linearity occur in ssp119 and abrupt-4xCO2 (as in \emph{Overshoot} and \emph{Abrupt}). This is consistent with the expectations from theory: under exponential forcing, we showed that regional warming is set by a single time scale and that the global-to-regional relationships should be perfectly linear, unless nonlinear feedbacks play a role. For abrupt-4xCO2, deviations from linearity are expected from theory and are related to the different heat capacities in the system. Deviations from linearity in abrupt-4xCO2 also have similar properties as in the idealized results (Figure \ref{fig:transientevaluations}), especially for the Southern Ocean pattern that shows a clear upward concavity and for Land that shows a downward concavity in the linear regression. For ssp119, the hysteresis behavior found in the idealized theory is not necessarily evident in the CMIP6 data. This is likely because temperatures in ssp119 do not decrease as rapidly as in \emph{Overshoot}, and the system does not have sufficient time to evolve after global temperature peaks in ssp119 to show the hysteresis. A few individual models show the hysteresis behavior for some regions (not shown), although this is not a widespread result across models. 

Quantitatively, CMIP6 data and the idealized results differ because we did not attempt to match the parameters of the idealized model to CMIP6 data. The \emph{Exponential}, \emph{Abrupt} and \emph{Overshoot} idealized models also lack any changes in heat transport, represent the ocean as a slab of water and have other important assumptions that make it quantitatively different from CMIP6 experiments. Yet, they seem to contain the necessary ingredients to explain most of the qualitative behavior of the warming pattern in these experiments.
Finally, while we limited our discussion to spatial averages over large regions (Land, Tropical Ocean and the Southern Ocean), similar qualitative relationships are also found for regional averages as well as locally for individual grid points \citep{lutjens_impact_2024}.

\subsection{Linear and nonlinear feedbacks (ii)}
\begin{figure}[b]
  \centerline{\includegraphics[width=19pc]{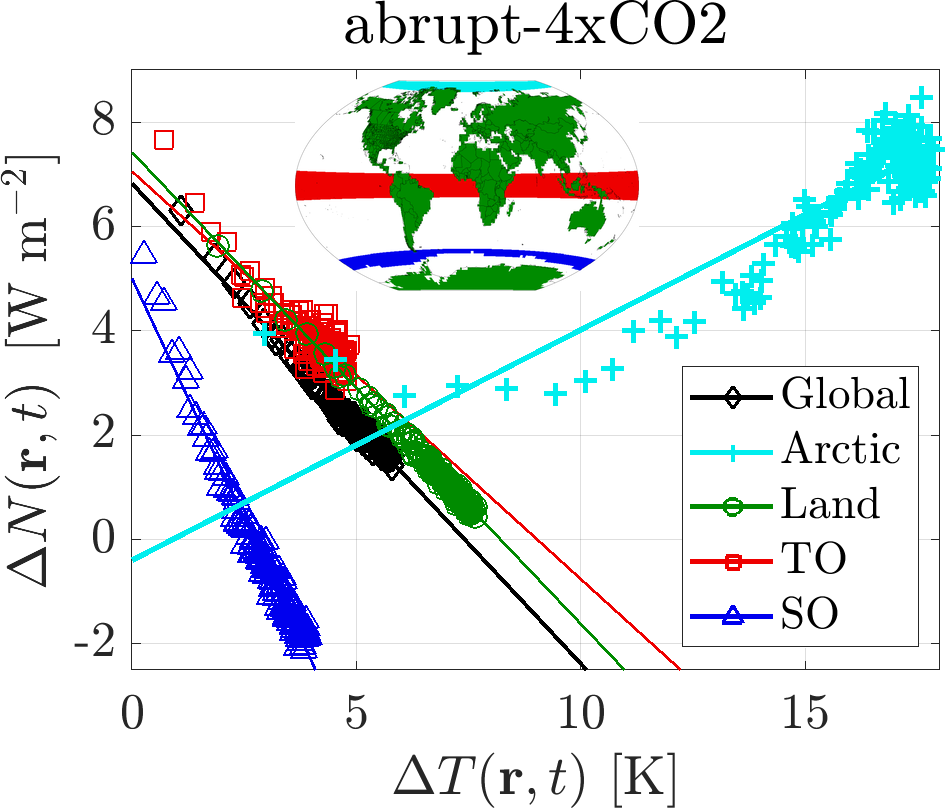}}
  \caption{Regional regression of TOA energy imbalance anomalies $\Delta N(\mathbf{r,t})$ and surface temperature anomalies $\Delta T(\mathbf{r,t})$ from CMIP6 abrupt-4xCO2 multi-model output.}
  \label{fig:cmipFD}
\end{figure}

For all the idealized cases that we presented (except for \emph{Albedo}), we assumed that the local TOA energy imbalance following a radiative perturbation is described by a linear function of surface temperature, where the slope is the constant feedback parameter. Are local feedbacks also constant in CMIP6 data? Figure \ref{fig:cmipFD} presents some evidence to argue that the constant feedback assumption is reasonably met for Land, TO and SO, whereas it does not hold for the Arctic. Specifically, Figure \ref{fig:cmipFD} shows TOA energy imbalance anomalies $\Delta N(\mathbf{r},t)$ as a function of surface temperature anomalies $\Delta T(\mathbf{r},t)$ in the abrupt-4xCO2 experiment. This is the typical \citet{gregory_new_2004} regression method to estimate forcing and feedbacks, applied at a local scale such as in \cite{armour_time-varying_2013}. 

For Land, TO and SO, the linear assumption holds well except for some nonlinearity early in the simulations. The regression $R^2$ are 0.99, 0.78, 0.97 and the regression slopes $\lambda(\mathbf{r})$, calculated over the entire time period, are -0.9, -0.8 and -1.8 W m$^{-2}$ K$^{-1}$, for Land, TO and SO, respectively. The initial nonlinearity can be understood in terms of fast and slow responses \citep{dong_intermodel_2020}. In the Arctic, $\Delta N$ is not a linear function of $\Delta T$, and is better described by a higher-order polynomial. As we showed in Section \ref{sec:idealized}\ref{subsec:nonlinearfeedbacks}, a nonlinear feedback is sufficient to violate the time-invariance of the regional warming pattern, which is likely the cause of the behavior observed in Figure \ref{fig:cmipPattern}. On the other hand, the near-linearity in feedbacks found for Land, TO and SO adds a further element to explain why the pattern is rather constant over these three regions in SSP scenarios. Furthermore, the shape of the $\Delta N(\Delta T)$ function is consistent with the physical mechanism in \emph{Albedo}. At the start of the simulation (1-5 years), radiative feedbacks decrease the initial imbalance imposed by quadrupling CO$_2$ instantaneously, consistently with a stabilizing negative feedback (equivalent to the fixed feedback that we impose in Section \ref{sec:idealized}\ref{subsec:nonlinearfeedbacks}). As the simulation progresses, the energy imbalance at TOA over the Arctic increases with increasing temperatures (positive feedback). This is likely associated with the \emph{Albedo} feedback described in Section \ref{sec:idealized}\ref{subsec:nonlinearfeedbacks}, where melting sea ice decreases the Arctic albedo and triggers a positive feedback that dominates over other stabilizing feedbacks. We do not rule out the possibility that additional mechanisms beyond the sea-ice albedo feedback, such as a nonlinear cloud responses, could contribute to the observed functional shape in CMIP6 simulations over the Arctic. An extendend pattern scaling approach, where the local-to-global dependence is a nonlinear function, seems a promising avenue to overcome the pattern scaling mismatch related to the nonlinear Arctic feedback.

\subsection{Invariant pattern of Forcing (iii)}

We now examine if the assumption of an invariant forcing pattern (iii) is sensible in CMIP6 simulations. We infer the forcing data $R(\mathbf{r},t)$ from the Radiative Forcing Model Intercomparison Project (RFMIP, \cite{pincus_radiative_2016}). $R(\mathbf{r},t)$ can be diagnosed from pairs of fixed-SST (sea surface temperature) integrations as the difference between TOA fluxes in the perturbed simulation and the control simulation \citep{pincus_radiative_2016}. As for the ScenarioMIP and DECK simulations, we present results as multi-model averages. Only a subset of five models ran all the experiments needed for this analysis (CanESM5, GFDL-CM4, HadGEM3-GC31-LL, IPSL-CM6A-LR, MIROC6). We analyze single forcing experiments (piClim-histghg, piClim-histaer and piClim-4xCO2) to disentangle the forcing patterns related to different agents, as well as their relative importance. In RFMIP, the piClim-histghg and piClim-histaer experiments have a transient forcing that match the historical and ssp245 experiments in ScenarioMIP for GHG and aerosols, respectively. No other SSP scenarios are available from RFMIP, and therefore we limit our analysis to ssp245. We also consider piClim-4xCO2, which is a time-slice experiment where forcing is held constant at 4 times the preindustrial CO$_2$ level.

\begin{figure}[bthp]
  \centerline{\includegraphics[width=19pc]{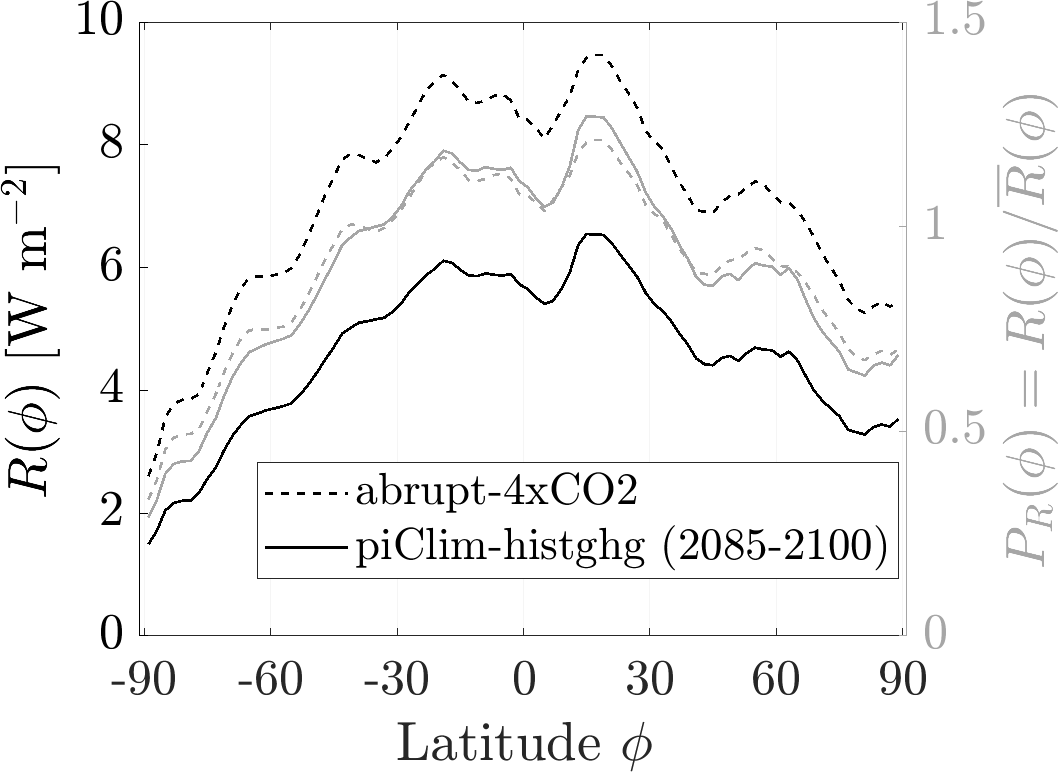}}
  \caption{Meridional structure of GHG and CO$_2$ forcing derived from RFMIP fixed-SST integrations. The left axis indicates values of the actual forcing $R(\phi)$; the right axis indicates values of the forcing pattern $P_R(\phi)=R(\phi)/\overline{R}(\phi)$.}
  \label{fig:cmipFR}
\end{figure}

\begin{figure*}[h]
\centerline{\includegraphics[width=0.8\textwidth]{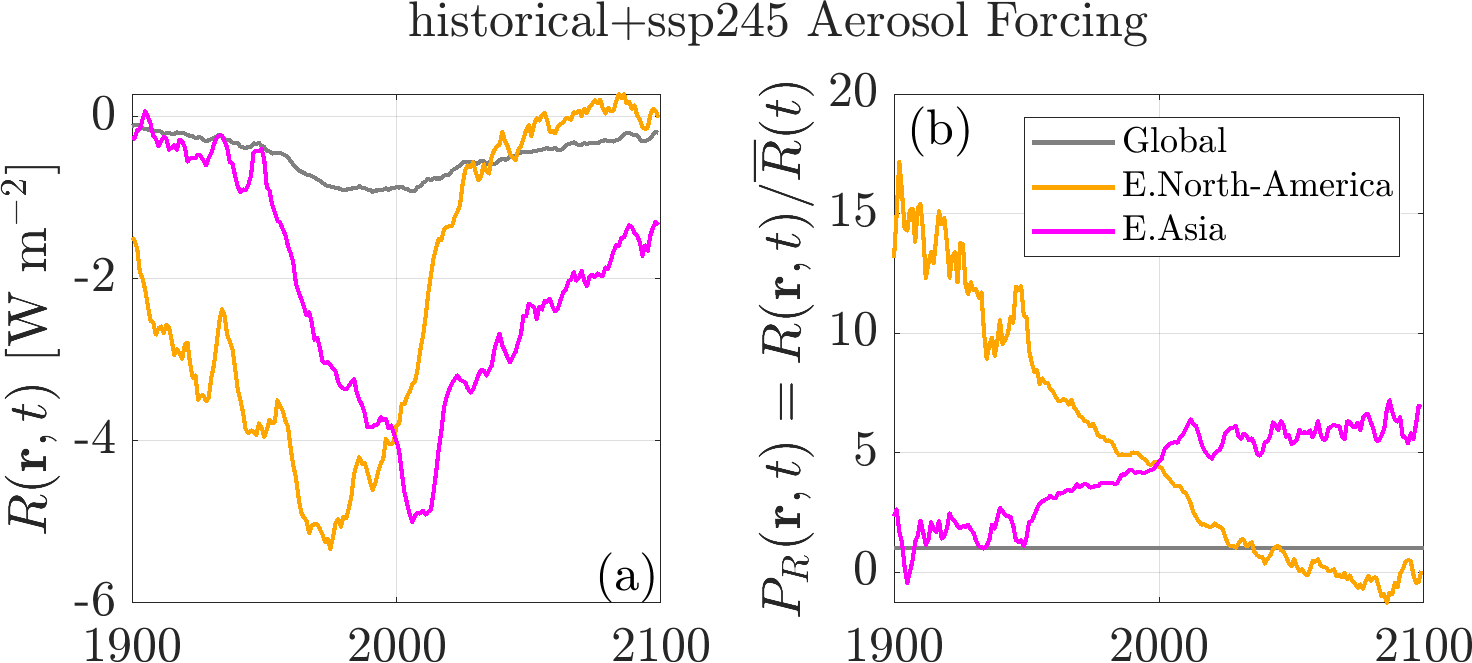}}
  \caption{Time evolution of (a) aerosol forcing $R(\mathbf{r})$ and (b) aerosol forcing pattern $P_R=R(\mathbf{r})/\overline{R}(\mathbf{r})$ over two IPCC regions, E.Asia and E.North-America, during the historical period and ssp245 projections. Data are derived from piClim-histaer and are smoothed with a 5-year moving average.}
  \label{fig:cmipFRCU}
\end{figure*}

Figure \ref{fig:cmipFR} shows the meridional dependence of well-mixed GHG forcing at the end of piClim-histghg (2085-2100), the CO$_2$ forcing from piClim-4xCO2 as well as their pattern $P_R(\phi)$, on the right axis. 
As expected, the magnitude of the forcing from piClim-4xCO2 and the end of piClim-histghg is very different, but the pattern is roughly constant. This is because well mixed gases have a clear meridional forcing structure that depend on the climatological lapse rate, water vapor and high clouds \citep{merlis_direct_2015}, but that is largely independent of the average forcing. 
This is not the case for short-lived aerosol forcing, as exemplified in Figure \ref{fig:cmipFRCU}, where we show the time evolution of $R(\mathbf{r},t)$ and $P_R(\mathbf{r},t)$ over two IPCC regions (East Asia, EAS, and East North-America, ENA; see \cite{iturbide_update_2020} for the regions definitions) in piClim-histaer. The forcing is negative owing to the scattering properties of the aerosol mixture and aerosol-cloud interactions, and resemble qualitatively the \emph{AerosolForcing} idealized case discussed in Section \ref{sec:idealized}\ref{subsec:forcingpattern}. The forcing over ENA peaks early (around 1970) and declines rapidly after that. Over EAS, the peak occurs decades later (around 2010) and declines more slowly. This has a large effect on the forcing patterns: over ENA, forcing is approximately 15 times the global average in 1900 and scales as the global average in 2100; over EAS, $R(\mathbf{r})$ is similar to the average in 1900 but is approximately 6 times the global average by 2100. 

\begin{figure*}[t]
\centerline{\includegraphics[width=1.0\textwidth]{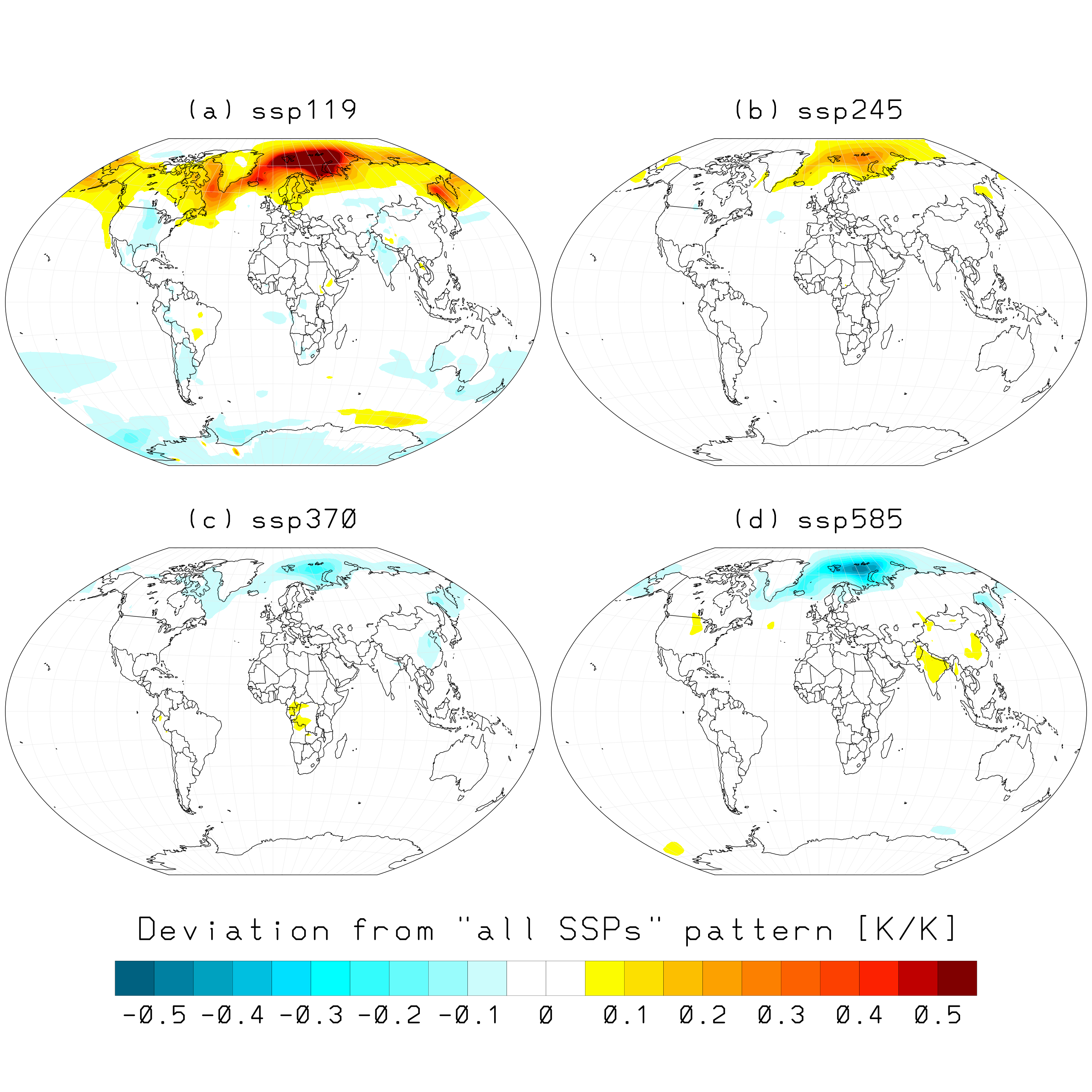}}
  \caption{Difference between the warming pattern calculated in individual SSP scenarios via global-to-local regression, and the "all SSPs" pattern calculated via the global-to-local regression including data from all SSPs simultaneously.}
  \label{fig:cmiperrors}
\end{figure*}

In realistic SSPs projections where aerosol forcing is superimposed to GHG forcing,  this effect is expected to matter considerably where the magnitude of the aerosol forcing is comparable to that of GHG forcing. This is definitely the case for regions like ENA and EAS, where the aerosol peak forcing ($\approx -5$ W m$^{-2}$) is comparable to GHG forcing in most SSP projections. Furthermore, different SSP projections vary considerably in their regional aerosol emissions, which also create multiple forcing (and warming) patterns. For regions where the peak aerosol forcing is small, the evolving forcing pattern mechanism gets masked by the much larger (and largely pattern-invariant) GHG forcing, which explains why the pattern averaged over land, for instance, is rather constant.  This is further illustrated in Figure \ref{fig:cmiperrors}, which shows the difference between the warming pattern calculated from regressing local temperatures against global temperatures using individual SSPs and the pattern calculated by including all SSPs in the regression. 
If the pattern is truly scenario-invariant, then the difference between between the individual patterns and the "all SSPs" pattern should be zero. Figure \ref{fig:cmiperrors} shows that is indeed the case for most of the planet (which is another confirmation of what we noted in Figure \ref{fig:cmipPattern}), with considerable differences only found in the Arctic (due to nonlinear feedbacks) and to some extent over South and South-East Asia, likely due to different aerosol projections in different SSPs.

\subsection{Dynamics (iv)}

Our idealized analysis uses one of the simplest approximation of changes in MHT and is intended to provide a zeroth-order estimate of what $\Delta \left( \nabla \cdot \mathbf{F}(\mathbf{r},t)\right) \ne 0$ can entail for evolving warming patterns. Many studies have sought to understand how MHT changes under warming scenarios in CMIP6 (e.g., \cite{cox_trends_2024,mecking_decrease_2023,fajber_atmospheric_2023,donohoe_partitioning_2020}), and suggested that MHT anomalies in coupled models may be more nuanced than a simple downgradient temperature diffusion \citep{armour_meridional_2019}. The total MHT (oceanic+atmospheric) has been shown to be nearly climate-state invariant from the last glacial maximum to experiments where CO$_2$ is increased fourfold \citep{donohoe_partitioning_2020}, with a compensating increase and decrease in poleward atmospheric and oceanic heat transport, respectively. Simulating compensating changes in our simple local energy balance framework would require a more sophisticated approach where the ocean and the atmosphere are modeled separately along with their coupling via surface fluxes (e.g., \cite{rose_ocean_2009}). In addition, heat transport changes depend on both temperature and velocity anomalies because of advection and eddy transport, introducing another layer of complexity that is not addressed here. Our analysis is also limited to meridional heat transport; additional complexity would arise without zonal averaging, and that complexity may be needed to capture phenomena such as the North Atlantic Warming Hole. A detailed investigation into how changes in heat transport influence warming patterns seems a promising avenue for further research. Different levels of idealization could be beneficial, from the one presented here to more complicated approaches that split oceanic and atmospheric heat transport, diffuse moist static energy \citep{rose_dependence_2014,armour_meridional_2019}, or account for changes in heat transport by parameterizing CMIP6 data.

\section{Discussion and Conclusions}

\label{sec:conclusions}

We used local energy balance arguments to show that the warming pattern in climate models is invariant under exponential and fixed-pattern forcing, linear feedbacks, and if dynamics can be represented as a linear and time-independent operator. We find that these conditions are roughly met for most CMIP6 SSP projections, which explains the widespread and successful use of pattern scaling for the past three decades in climate impact assessments. Notable exceptions are the Arctic region, where nonlinear feedbacks play a major role, and regions with strong scenario-dependent aerosol forcing. This is exemplified in Figure \ref{fig:cmiperrors}, where considerable differences in warming patterns are found where we expect a large influence from nonlinear feedbacks and aerosols. For abrupt changes (or forcing that strongly deviate from an exponential, such as overshoot scenarios), the warming pattern is not invariant in the transient adjustment as different time scales in the climate system emerge, even if feedbacks are linear and if the forcing pattern does not vary. 

We explore some implications of our findings. First, our results demonstrate that pattern scaling is a robust approximation for temperature in SSP projections, and provide some fundamental understanding on why more complicated approaches, such as machine learning and kernel emulators, have provided marginal improvements in accuracy over simple pattern scaling calculations \citep{watsonparris_climatebench_2022,freese_spatially_2024,womack2024rapid,lutjens_impact_2024}. For the SSP projections, room for progress over pattern scaling is mostly confined to the Arctic and to regions with large aerosol forcing. The theory also provides insights into why previous work found the largest pattern scaling errors in overshoot scenarios and abrupt experiments \citep{wells_understanding_2023,womack2024rapid}, such as idealized geoengineering scenarios \citep{schaller_asymmetry_2014}.

Second, these findings help reconcile pattern scaling, which assumes that $\Delta T(\mathbf{r},t)/\overline{\Delta T}(t)$ is constant, with the pattern effect, which explains time-varying effective climate sensitivity with time-varying $\Delta T(\mathbf{r},t)/\overline{\Delta T}(t)$. The two lines of research are not necessarily contradictory, and the difference between \emph{pattern scaling} and the \emph{pattern effect} is conceptually captured by the top and middle rows in Figure \ref{fig:transientevaluations}. It can also be understood in terms of the fast and slow components of warming introduced by \cite{held_probing_2010} and \cite{geoffroy_transient_2013}. In typical end-of-century scenarios where forcing increases, pattern scaling is a good approximation because the fast component associated with the atmosphere and the upper ocean reacts to the increasing forcing. On longer time scales and for constant forcing, the slow component emerges and becomes comparable to the fast component, changing the effective climate sensitivity (i.e., the pattern effect).  Another separate nuance is that \emph{pattern scaling} is typically used to calculate the forced changes with ensemble averages, whereas the \emph{pattern effect} literature has also focused on observed internal variability in the warming pattern to interpret observational constraints on climate sensitivity \citep{dessler_potential_2020,armour_sea-surface_2024}.

Finally, while we have provided additional understanding on the mechanisms for the time evolution (or lack thereof) of the warming pattern, we do not yet have a comprehensive theory on what sets the warming patterns in climate models. Using the TOA energy balance, we have outlined some relevant parameters (such as the forcing pattern and local feedbacks, Equation \ref{eq:transientpatternEXP}); however, other important parameters, such as the surface partitioning between latent and sensible heat, cannot be fully captured within our TOA idealization. We speculate that a potential avenue for a more comprehensive theory should rest on the surface energy balance and boundary layer modeling \citep{cronin_sensitivity_2013}. A comprehensive theory that explains what sets the warming pattern in climate models could help interpret the intermodel spread in regional warming and potentially the discrepancies between models and observations. In addition, the role of heat transport anomalies, which we highly simplified in this work, merits further exploration.

\acknowledgments We acknowledge support from Schmidt Sciences, LLC and the MIT Climate Grand Challenges through the Bringing Computation to the Climate Challenge (BC3) project. We also acknowledge the MIT \emph{Svante} cluster supported by the Center for Sustainability Science and Strategy for computing resources. We are grateful to the BC3 team for insightful discussions about this work.

%
%
\datastatement All data used in this work are publicly available through the World Climate Research Program (WCRP) Coupled Model Intercomparison Project 6 (CMIP6).

%


\appendix[A] 


\appendixtitle{Generalized Local Energy Balance and Nonlocal Radiative Effects}

In Section \ref{sec:idealized}, we assumed that the local energy imbalance at TOA is a linear function of local surface temperature. However, it has been shown that the local radiative response can also depend on nonlocal surface temperature responses via atmospheric dynamics \citep{feldl_nonlinear_2013,dong_attributing_2019,dong_intermodel_2020,blochjohnson_greens_2024}, and can be represented as a linear combination of surface temperature at all locations:
\begin{equation}
\Delta N(\mathbf{r},t) = R(\mathbf{r},t) + \mathbf{J}(\mathbf{r},\mathbf{r'})\Delta T(\mathbf{r'},t)
\end{equation}

Where $J(\mathbf{r},\mathbf{r'})$ is a matrix that quantifies the contribution to the local $\Delta N(\mathbf{r})$ from every location $\mathbf{r'}$. This is a more general representation of Equation \ref{eq:linearimbalance}, and is equivalent to Equation \ref{eq:linearimbalance} only if $J(\mathbf{r},\mathbf{r'})$ is diagonal with elements $\lambda(\mathbf{r})$. The matrix $\mathbf{J}$ is typically diagnosed via the Green's function approach \citep{dong_attributing_2019,blochjohnson_greens_2024}. With this representation of the TOA energy imbalance, the linear system that describes the temperature response to forcing (Equation \ref{eq:transient}) becomes coupled:
\begin{equation}
C(\mathbf{r})\frac{dT(\mathbf{r},t)}{dt} = R(\mathbf{r},t) + \mathbf{J(\mathbf{r},\mathbf{r'})} T(\mathbf{r'},t)
\label{eq:generalizedbalance}
\end{equation}
In Equation \ref{eq:generalizedbalance}, the temperature time evolution at any location $\mathbf{r}$ also depends on the temperature evolution at every other locations. However, Equation \ref{eq:generalizedbalance} also has exponential eigenfunctions (see \citet{raupach_exponential_2013} for the proof), and therefore the results described in Section \ref{sec:idealized}\ref{subsec:transient} are also valid when considering nonlocal radiative feedbacks. The key assumption is that the nonlocal effects are linearly additive, which has been shown to be a good first order approximation \citep{dong_attributing_2019,blochjohnson_greens_2024}. In other words, the warming pattern remains invariant under exponential forcing as long as the underlying system is linear, even if the system is coupled. We also discuss the same result in Section \ref{sec:idealized}\ref{subsec:dynamics}, where we couple the system by assuming diffusive dynamics.

\appendix[B] 


\appendixtitle{Transient Solution with Linear and Polynomial Forcing}

For completeness, we also discuss two additional cases where the forcing is either linear or polynomial. The solution of the transient energy balance with no diffusion, constant stabilizing feedbacks and invariant forcing pattern (Equation \ref{eq:transientTimeScale}), with initial condition $\Delta T(\mathbf{r},t=0) = 0$ is:

\begin{equation}
    \Delta T(\mathbf{r},t) = kf_R(\mathbf{r})\tau^2(\mathbf{r})\left(\frac{t}{\tau(\mathbf{r})} + \exp{\left(-\frac{t}{\tau(\mathbf{r})}\right) - 1}\right)
    \label{eq:linsolution}
\end{equation}

The asymptotic behavior of Equation \ref{eq:linsolution} for $t \to\infty$ is  $kf_R\tau t$, which means that temperatures at every location lock in with the forcing and grow linearly with time, with different coefficients set by $kf_R(\mathbf{r})\tau(\mathbf{r})$. This implies that, for large $t$, the pattern invariance is also found for the linear forcing case. For small $t$, the warming pattern varies in time because of the decaying exponential $\exp\left(-t/\tau\right)$ that arises from the homogeneous solution. We can approximate the system to be pattern invariant for $t \gg \tau$, where $t/\tau \gg \exp(-t/\tau)$. In other words, for $t \gg \tau$ the linear part of the solution $t/\tau$ is much greater than the exponential term $\exp\left(-t/\tau\right)$, and temperatures grow approximately linearly. 

The asymptotic warming pattern under linear forcing is $\Delta T(\mathbf{r})/\overline{\Delta T}=\overline{\lambda}_h/\lambda(\mathbf{r})$, where $\overline{(\cdot)}_h$ is the harmonic mean operator. This is the same asymptotic pattern as in the constant forcing case, but is different from the exponential forcing pattern which has a dependence on the forcing time scale $\tau_0$ (Equation \ref{eq:transientpatternEXP}). However, the warming pattern is also equal to $\overline{\lambda}_h/\lambda(\mathbf{r})$ in the exponential case if $\max{\left(\tau(\mathbf{r})\right)} \ll \tau_0$, so that the effect of $\tau_0$ on the warming pattern can be neglected (Equation \ref{eq:transientpatternEXP}). This is not the case for our three-region model example, where $\max{\left(\tau(\mathbf{r})\right)} = 287$ years and $\tau_0=50$ years, which explains why the top row patterns in Figure \ref{fig:transientevaluations} are different from the middle row. In reality, the condition $\tau(\mathbf{r}) \ll \tau_0$ is likely met for a large fraction of the planet given the fast response of the atmosphere and the upper ocean (on the order of a few years), with the exception of regions characterized by strong ocean heat uptake (such as the Southern Ocean).

Finally, similar arguments of temperatures \emph{locking-in} with the forcing can be made for polynomials, because the particular solutions to Equation \ref{eq:transient} are also polynomials of the same degree. After an initial transient where the exponential homogeneous solution decays and the system adjusts to the initial forcing imbalance, a dynamic equilibrium is reached where the pattern is invariant.

%



\bibliographystyle{ametsocV6}
\bibliography{references}

\begin{thebibliography}{72}
\providecommand{\natexlab}[1]{#1}
\providecommand{\url}[1]{\texttt{#1}}
\renewcommand{\UrlFont}{\rmfamily}
\providecommand{\urlprefix}{URL }
\expandafter\ifx\csname urlstyle\endcsname\relax
  \providecommand{\doi}[1]{https://doi.org/\discretionary{}{}{}#1}\else
  \providecommand{\doi}{https://doi.org/\discretionary{}{}{}\begingroup \urlstyle{rm}\Url}\fi
\providecommand{\eprint}[2][]{\url{#2}}

\bibitem[{Andrews et~al.(2015)Andrews, Gregory,, and Webb}]{andrews_dependence_2015}
Andrews, T., J.~M. Gregory, and M.~J. Webb, 2015: The {Dependence} of {Radiative} {Forcing} and {Feedback} on {Evolving} {Patterns} of {Surface} {Temperature} {Change} in {Climate} {Models}. \textit{Journal of Climate}, \textbf{28~(4)}, 1630--1648, \doi{10.1175/JCLI-D-14-00545.1}.

\bibitem[{Andrews et~al.(2018)}]{andrews_accounting_2018}
Andrews, T., and Coauthors, 2018: Accounting for {Changing} {Temperature} {Patterns} {Increases} {Historical} {Estimates} of {Climate} {Sensitivity}. \textit{Geophysical Research Letters}, \textbf{45~(16)}, 8490--8499, \doi{10.1029/2018GL078887}.

\bibitem[{Armour(2017)}]{armour_energy_2017}
Armour, K.~C., 2017: Energy budget constraints on climate sensitivity in light of inconstant climate feedbacks. \textit{Nature Climate Change}, \textbf{7~(5)}, 331--335, \doi{10.1038/nclimate3278}.

\bibitem[{Armour et~al.(2013)Armour, Bitz,, and Roe}]{armour_time-varying_2013}
Armour, K.~C., C.~M. Bitz, and G.~H. Roe, 2013: Time-{Varying} {Climate} {Sensitivity} from {Regional} {Feedbacks}. \textit{Journal of Climate}, \textbf{26~(13)}, 4518--4534, \doi{10.1175/JCLI-D-12-00544.1}.

\bibitem[{Armour et~al.(2016)Armour, Marshall, Scott, Donohoe,, and Newsom}]{armour_southern_2016}
Armour, K.~C., J.~Marshall, J.~R. Scott, A.~Donohoe, and E.~R. Newsom, 2016: Southern {Ocean} warming delayed by circumpolar upwelling and equatorward transport. \textit{Nature Geoscience}, \textbf{9~(7)}, 549--554, \doi{10.1038/ngeo2731}.

\bibitem[{Armour et~al.(2019)Armour, Siler, Donohoe,, and Roe}]{armour_meridional_2019}
Armour, K.~C., N.~Siler, A.~Donohoe, and G.~H. Roe, 2019: Meridional {Atmospheric} {Heat} {Transport} {Constrained} by {Energetics} and {Mediated} by {Large}-{Scale} {Diffusion}. \textit{Journal of Climate}, \textbf{32~(12)}, 3655--3680, \doi{10.1175/JCLI-D-18-0563.1}.

\bibitem[{Armour et~al.(2024)}]{armour_sea-surface_2024}
Armour, K.~C., and Coauthors, 2024: Sea-surface temperature pattern effects have slowed global warming and biased warming-based constraints on climate sensitivity. \textit{Proceedings of the National Academy of Sciences}, \textbf{121~(12)}, e2312093\,121, \doi{10.1073/pnas.2312093121}.

\bibitem[{Bates(2012)}]{bates_climate_2012}
Bates, J.~R., 2012: Climate stability and sensitivity in some simple conceptual models. \textit{Climate Dynamics}, \textbf{38~(3-4)}, 455--473, \doi{10.1007/s00382-010-0966-0}.

\bibitem[{Bauer et~al.(2022)Bauer, Tsigaridis, Faluvegi, Nazarenko, Miller, Kelley,, and Schmidt}]{bauer_turning_2022}
Bauer, S.~E., K.~Tsigaridis, G.~Faluvegi, L.~Nazarenko, R.~L. Miller, M.~Kelley, and G.~Schmidt, 2022: The {Turning} {Point} of the {Aerosol} {Era}. \textit{Journal of Advances in Modeling Earth Systems}, \textbf{14~(12)}, e2022MS003\,070, \doi{10.1029/2022MS003070}.

\bibitem[{Beusch et~al.(2020)Beusch, Gudmundsson,, and Seneviratne}]{beusch_emulating_2020}
Beusch, L., L.~Gudmundsson, and S.~I. Seneviratne, 2020: Emulating {Earth} system model temperatures with {MESMER}: {From} global mean temperature trajectories to grid-point-level realizations on land. \textit{Earth System Dynamics}, \textbf{11~(1)}, 139--159, \doi{10.5194/esd-11-139-2020}.

\bibitem[{Bloch‐Johnson et~al.(2024)}]{blochjohnson_greens_2024}
Bloch‐Johnson, J., and Coauthors, 2024: The {Green}'s {Function} {Model} {Intercomparison} {Project} ({GFMIP}) {Protocol}. \textit{Journal of Advances in Modeling Earth Systems}, \textbf{16~(2)}, e2023MS003\,700, \doi{10.1029/2023MS003700}.

\bibitem[{Bronselaer and Zanna(2020)Bronselaer, and Zanna}]{bronselaer_heat_2020}
Bronselaer, B., and L.~Zanna, 2020: Heat and carbon coupling reveals ocean warming due to circulation changes. \textit{Nature}, \textbf{584~(7820)}, 227--233, \doi{10.1038/s41586-020-2573-5}.

\bibitem[{Byrne and O’Gorman(2018)Byrne, and O’Gorman}]{byrne_trends_2018}
Byrne, M.~P., and P.~A. O’Gorman, 2018: Trends in continental temperature and humidity directly linked to ocean warming. \textit{Proceedings of the National Academy of Sciences}, \textbf{115~(19)}, 4863--4868, \doi{10.1073/pnas.1722312115}.

\bibitem[{Callahan and Mankin(2022)Callahan, and Mankin}]{callahan_national_2022}
Callahan, C.~W., and J.~S. Mankin, 2022: National attribution of historical climate damages. \textit{Climatic Change}, \textbf{172~(3-4)}, 40, \doi{10.1007/s10584-022-03387-y}.

\bibitem[{Cox et~al.(2024)Cox, Donohoe, Armour, Frierson,, and Roe}]{cox_trends_2024}
Cox, T., A.~Donohoe, K.~C. Armour, D.~M.~W. Frierson, and G.~H. Roe, 2024: Trends in {Atmospheric} {Heat} {Transport} {Since} 1980. \textit{Journal of Climate}, \textbf{37~(5)}, 1539--1550, \doi{10.1175/JCLI-D-23-0385.1}.

\bibitem[{Cronin(2013)}]{cronin_sensitivity_2013}
Cronin, T.~W., 2013: A sensitivity theory for the equilibrium boundary layer over land. \textit{Journal of Advances in Modeling Earth Systems}, \textbf{5~(4)}, 764--784, \doi{10.1002/jame.20048}.

\bibitem[{Dessler(2020)}]{dessler_potential_2020}
Dessler, A.~E., 2020: Potential {Problems} {Measuring} {Climate} {Sensitivity} from the {Historical} {Record}. \textit{Journal of Climate}, \textbf{33~(6)}, 2237--2248, \doi{10.1175/JCLI-D-19-0476.1}.

\bibitem[{Dong et~al.(2020)Dong, Armour, Zelinka, Proistosescu, Battisti, Zhou,, and Andrews}]{dong_intermodel_2020}
Dong, Y., K.~C. Armour, M.~D. Zelinka, C.~Proistosescu, D.~S. Battisti, C.~Zhou, and T.~Andrews, 2020: Intermodel {Spread} in the {Pattern} {Effect} and {Its} {Contribution} to {Climate} {Sensitivity} in {CMIP5} and {CMIP6} {Models}. \textit{Journal of Climate}, \textbf{33~(18)}, 7755--7775, \doi{10.1175/JCLI-D-19-1011.1}.

\bibitem[{Dong et~al.(2019)Dong, Proistosescu, Armour,, and Battisti}]{dong_attributing_2019}
Dong, Y., C.~Proistosescu, K.~C. Armour, and D.~S. Battisti, 2019: Attributing {Historical} and {Future} {Evolution} of {Radiative} {Feedbacks} to {Regional} {Warming} {Patterns} using a {Green}’s {Function} {Approach}: {The} {Preeminence} of the {Western} {Pacific}. \textit{Journal of Climate}, \textbf{32~(17)}, 5471--5491, \doi{10.1175/JCLI-D-18-0843.1}.

\bibitem[{Donohoe et~al.(2020)Donohoe, Armour, Roe, Battisti,, and Hahn}]{donohoe_partitioning_2020}
Donohoe, A., K.~C. Armour, G.~H. Roe, D.~S. Battisti, and L.~Hahn, 2020: The {Partitioning} of {Meridional} {Heat} {Transport} from the {Last} {Glacial} {Maximum} to {CO2} {Quadrupling} in {Coupled} {Climate} {Models}. \textit{Journal of Climate}, \textbf{33~(10)}, 4141--4165, \doi{10.1175/JCLI-D-19-0797.1}.

\bibitem[{Estrada et~al.(2017)Estrada, Botzen,, and Tol}]{estrada_global_2017}
Estrada, F., W.~J.~W. Botzen, and R.~S.~J. Tol, 2017: A global economic assessment of city policies to reduce climate change impacts. \textit{Nature Climate Change}, \textbf{7~(6)}, 403--406, \doi{10.1038/nclimate3301}.

\bibitem[{Eyring et~al.(2016)Eyring, Bony, Meehl, Senior, Stevens, Stouffer,, and Taylor}]{eyring_overview_2016}
Eyring, V., S.~Bony, G.~A. Meehl, C.~A. Senior, B.~Stevens, R.~J. Stouffer, and K.~E. Taylor, 2016: Overview of the {Coupled} {Model} {Intercomparison} {Project} {Phase} 6 ({CMIP6}) experimental design and organization. \textit{Geoscientific Model Development}, \textbf{9~(5)}, 1937--1958, \doi{10.5194/gmd-9-1937-2016}.

\bibitem[{Fajber et~al.(2023)Fajber, Donohoe, Ragen, Armour,, and Kushner}]{fajber_atmospheric_2023}
Fajber, R., A.~Donohoe, S.~Ragen, K.~C. Armour, and P.~J. Kushner, 2023: Atmospheric heat transport is governed by meridional gradients in surface evaporation in modern-day earth-like climates. \textit{Proceedings of the National Academy of Sciences}, \textbf{120~(25)}, e2217202\,120, \doi{10.1073/pnas.2217202120}.

\bibitem[{Feldl and Roe(2013)Feldl, and Roe}]{feldl_nonlinear_2013}
Feldl, N., and G.~H. Roe, 2013: The {Nonlinear} and {Nonlocal} {Nature} of {Climate} {Feedbacks}. \textit{Journal of Climate}, \textbf{26~(21)}, 8289--8304, \doi{10.1175/JCLI-D-12-00631.1}.

\bibitem[{Freese et~al.(2024)Freese, Giani, Fiore,, and Selin}]{freese_spatially_2024}
Freese, L.~M., P.~Giani, A.~M. Fiore, and N.~E. Selin, 2024: Spatially {Resolved} {Temperature} {Response} {Functions} to {CO} $_{\textrm{2}}$ {Emissions}. \textit{Geophysical Research Letters}, \textbf{51~(15)}, e2024GL108\,788, \doi{10.1029/2024GL108788}.

\bibitem[{Geoffroy et~al.(2013)Geoffroy, Saint-Martin, Bellon, Voldoire, Olivié,, and Tytéca}]{geoffroy_transient_2013}
Geoffroy, O., D.~Saint-Martin, G.~Bellon, A.~Voldoire, D.~J.~L. Olivié, and S.~Tytéca, 2013: Transient {Climate} {Response} in a {Two}-{Layer} {Energy}-{Balance} {Model}. {Part} {II}: {Representation} of the {Efficacy} of {Deep}-{Ocean} {Heat} {Uptake} and {Validation} for {CMIP5} {AOGCMs}. \textit{Journal of Climate}, \textbf{26~(6)}, 1859--1876, \doi{10.1175/JCLI-D-12-00196.1}.

\bibitem[{Gillett(2023)}]{gillett_warming_2023}
Gillett, N.~P., 2023: Warming proportional to cumulative carbon emissions not explained by heat and carbon sharing mixing processes. \textit{Nature Communications}, \textbf{14~(1)}, 6466, \doi{10.1038/s41467-023-42111-x}.

\bibitem[{Gregory et~al.(2004)}]{gregory_new_2004}
Gregory, J.~M., and Coauthors, 2004: A new method for diagnosing radiative forcing and climate sensitivity. \textit{Geophysical Research Letters}, \textbf{31~(3)}, 2003GL018\,747, \doi{10.1029/2003GL018747}.

\bibitem[{Held et~al.(1981)Held, Linder,, and Suarez}]{held_albedo_1981}
Held, I.~M., D.~I. Linder, and M.~J. Suarez, 1981: Albedo {Feedback}, the {Meridional} {Structure} of the {Effective} {Heat} {Diffusivity}, and {Climatic} {Sensitivity}: {Results} from {Dynamic} and {Diffusive} {Models}. \textit{Journal of the Atmospheric Sciences}, \textbf{38~(9)}, 1911--1927, \doi{10.1175/1520-0469(1981)038<1911:AFTMSO>2.0.CO;2}.

\bibitem[{Held and Suarez(1974)Held, and Suarez}]{held_simple_1974}
Held, I.~M., and M.~J. Suarez, 1974: Simple albedo feedback models of the icecaps. \textit{Tellus}, \textbf{26~(6)}, 613--629, \doi{10.1111/j.2153-3490.1974.tb01641.x}.

\bibitem[{Held et~al.(2010)Held, Winton, Takahashi, Delworth, Zeng,, and Vallis}]{held_probing_2010}
Held, I.~M., M.~Winton, K.~Takahashi, T.~Delworth, F.~Zeng, and G.~K. Vallis, 2010: Probing the {Fast} and {Slow} {Components} of {Global} {Warming} by {Returning} {Abruptly} to {Preindustrial} {Forcing}. \textit{Journal of Climate}, \textbf{23~(9)}, 2418--2427, \doi{10.1175/2009JCLI3466.1}.

\bibitem[{Hsiang et~al.(2017)}]{hsiang_estimating_2017}
Hsiang, S., and Coauthors, 2017: Estimating economic damage from climate change in the {United} {States}. \textit{Science}, \textbf{356~(6345)}, 1362--1369, \doi{10.1126/science.aal4369}.

\bibitem[{Iturbide et~al.(2020)}]{iturbide_update_2020}
Iturbide, M., and Coauthors, 2020: An update of {IPCC} climate reference regions for subcontinental analysis of climate model data: definition and aggregated datasets. \textit{Earth System Science Data}, \textbf{12~(4)}, 2959--2970, \doi{10.5194/essd-12-2959-2020}.

\bibitem[{Joshi et~al.(2008)Joshi, Gregory, Webb, Sexton,, and Johns}]{joshi_mechanisms_2008}
Joshi, M.~M., J.~M. Gregory, M.~J. Webb, D.~M.~H. Sexton, and T.~C. Johns, 2008: Mechanisms for the land/sea warming contrast exhibited by simulations of climate change. \textit{Climate Dynamics}, \textbf{30~(5)}, 455--465, \doi{10.1007/s00382-007-0306-1}.

\bibitem[{Keil et~al.(2020)Keil, Mauritsen, Jungclaus, Hedemann, Olonscheck,, and Ghosh}]{keil_multiple_2020}
Keil, P., T.~Mauritsen, J.~Jungclaus, C.~Hedemann, D.~Olonscheck, and R.~Ghosh, 2020: Multiple drivers of the {North} {Atlantic} warming hole. \textit{Nature Climate Change}, \textbf{10~(7)}, 667--671, \doi{10.1038/s41558-020-0819-8}.

\bibitem[{Knutti et~al.(2017)Knutti, Rugenstein,, and Hegerl}]{knutti_beyond_2017}
Knutti, R., M.~A.~A. Rugenstein, and G.~C. Hegerl, 2017: Beyond equilibrium climate sensitivity. \textit{Nature Geoscience}, \textbf{10~(10)}, 727--736, \doi{10.1038/ngeo3017}.

\bibitem[{Leduc et~al.(2016)Leduc, Matthews,, and De~Elía}]{leduc_regional_2016}
Leduc, M., H.~D. Matthews, and R.~De~Elía, 2016: Regional estimates of the transient climate response to cumulative {CO2} emissions. \textit{Nature Climate Change}, \textbf{6~(5)}, 474--478, \doi{10.1038/nclimate2913}.

\bibitem[{Lütjens et~al.(2024)Lütjens, Ferrari, Watson-Parris,, and Selin}]{lutjens_impact_2024}
Lütjens, B., R.~Ferrari, D.~Watson-Parris, and N.~Selin, 2024: The impact of internal variability on benchmarking deep learning climate emulators. arXiv, \urlprefix\url{http://arxiv.org/abs/2408.05288}, arXiv:2408.05288 [cs].

\bibitem[{MacDougall(2016)}]{macdougall_transient_2016}
MacDougall, A.~H., 2016: The {Transient} {Response} to {Cumulative} {CO2} {Emissions}: a {Review}. \textit{Current Climate Change Reports}, \textbf{2~(1)}, 39--47, \doi{10.1007/s40641-015-0030-6}.

\bibitem[{MacDougall(2017)}]{macdougall_oceanic_2017}
MacDougall, A.~H., 2017: The oceanic origin of path-independent carbon budgets. \textit{Scientific Reports}, \textbf{7~(1)}, 10\,373, \doi{10.1038/s41598-017-10557-x}.

\bibitem[{Matthews and Caldeira(2008)Matthews, and Caldeira}]{matthews_stabilizing_2008}
Matthews, H.~D., and K.~Caldeira, 2008: Stabilizing climate requires near‐zero emissions. \textit{Geophysical Research Letters}, \textbf{35~(4)}, 2007GL032\,388, \doi{10.1029/2007GL032388}.

\bibitem[{Matthews et~al.(2009)Matthews, Gillett, Stott,, and Zickfeld}]{matthews_proportionality_2009}
Matthews, H.~D., N.~P. Gillett, P.~A. Stott, and K.~Zickfeld, 2009: The proportionality of global warming to cumulative carbon emissions. \textit{Nature}, \textbf{459~(7248)}, 829--832, \doi{10.1038/nature08047}.

\bibitem[{Matthews et~al.(2020)}]{matthews_opportunities_2020}
Matthews, H.~D., and Coauthors, 2020: Opportunities and challenges in using remaining carbon budgets to guide climate policy. \textit{Nature Geoscience}, \textbf{13~(12)}, 769--779, \doi{10.1038/s41561-020-00663-3}.

\bibitem[{Mecking and Drijfhout(2023)Mecking, and Drijfhout}]{mecking_decrease_2023}
Mecking, J.~V., and S.~S. Drijfhout, 2023: The decrease in ocean heat transport in response to global warming. \textit{Nature Climate Change}, \textbf{13~(11)}, 1229--1236, \doi{10.1038/s41558-023-01829-8}.

\bibitem[{Meinshausen et~al.(2020)}]{meinshausen_shared_2020}
Meinshausen, M., and Coauthors, 2020: The shared socio-economic pathway ({SSP}) greenhouse gas concentrations and their extensions to 2500. \textit{Geoscientific Model Development}, \textbf{13~(8)}, 3571--3605, \doi{10.5194/gmd-13-3571-2020}.

\bibitem[{Merlis(2014)}]{merlis_interacting_2014}
Merlis, T.~M., 2014: Interacting components of the top‐of‐atmosphere energy balance affect changes in regional surface temperature. \textit{Geophysical Research Letters}, \textbf{41~(20)}, 7291--7297, \doi{10.1002/2014GL061700}.

\bibitem[{Merlis(2015)}]{merlis_direct_2015}
Merlis, T.~M., 2015: Direct weakening of tropical circulations from masked {CO} $_{\textrm{2}}$ radiative forcing. \textit{Proceedings of the National Academy of Sciences}, \textbf{112~(43)}, 13\,167--13\,171, \doi{10.1073/pnas.1508268112}.

\bibitem[{Mitchell(2003)}]{mitchell_pattern_2003}
Mitchell, T.~D., 2003: Pattern {Scaling}: {An} {Examination} of the {Accuracy} of the {Technique} for {Describing} {Future} {Climates}. \textit{Climatic Change}, \textbf{60}, 217--242.

\bibitem[{North(1975)}]{north_analytical_1975}
North, G.~R., 1975: Analytical {Solution} to a {Simple} {Climate} {Model} with {Diffusive} {Heat} {Transport}. \textit{Journal of the Atmospheric Sciences}, \textbf{32~(7)}, 1301--1307, \doi{10.1175/1520-0469(1975)032<1301:ASTASC>2.0.CO;2}.

\bibitem[{O'Neill et~al.(2016)}]{oneill_scenario_2016}
O'Neill, B.~C., and Coauthors, 2016: The {Scenario} {Model} {Intercomparison} {Project} ({ScenarioMIP}) for {CMIP6}. \textit{Geoscientific Model Development}, \textbf{9~(9)}, 3461--3482, \doi{10.5194/gmd-9-3461-2016}.

\bibitem[{Osborn et~al.(2018)Osborn, Wallace, Lowe,, and Bernie}]{osborn_performance_2018}
Osborn, T.~J., C.~J. Wallace, J.~A. Lowe, and D.~Bernie, 2018: Performance of {Pattern}-{Scaled} {Climate} {Projections} under {High}-{End} {Warming}. {Part} {I}: {Surface} {Air} {Temperature} over {Land}. \textit{Journal of Climate}, \textbf{31}, 5667--5680, \doi{10.1175/JCLI}.

\bibitem[{Pfister and Stocker(2021)Pfister, and Stocker}]{pfister_changes_2021}
Pfister, P.~L., and T.~F. Stocker, 2021: Changes in {Local} and {Global} {Climate} {Feedbacks} in the {Absence} of {Interactive} {Clouds}: {Southern} {Ocean}–{Climate} {Interactions} in {Two} {Intermediate}-{Complexity} {Models}. \textit{Journal of Climate}, \textbf{34~(2)}, 755--772, \doi{10.1175/JCLI-D-20-0113.1}.

\bibitem[{Pincus et~al.(2016)Pincus, Forster,, and Stevens}]{pincus_radiative_2016}
Pincus, R., P.~M. Forster, and B.~Stevens, 2016: The {Radiative} {Forcing} {Model} {Intercomparison} {Project} ({RFMIP}): experimental protocol for {CMIP6}. \textit{Geoscientific Model Development}, \textbf{9~(9)}, 3447--3460, \doi{10.5194/gmd-9-3447-2016}.

\bibitem[{Previdi et~al.(2021)Previdi, Smith,, and Polvani}]{previdi_arctic_2021}
Previdi, M., K.~L. Smith, and L.~M. Polvani, 2021: Arctic amplification of climate change: a review of underlying mechanisms. \textit{Environmental Research Letters}, \textbf{16~(9)}, 093\,003, \doi{10.1088/1748-9326/ac1c29}.

\bibitem[{Raupach(2013)}]{raupach_exponential_2013}
Raupach, M.~R., 2013: The exponential eigenmodes of the carbon-climate system, and their implications for ratios of responses to forcings. \textit{Earth System Dynamics}, \textbf{4~(1)}, 31--49, \doi{10.5194/esd-4-31-2013}.

\bibitem[{Rose et~al.(2014)Rose, Armour, Battisti, Feldl,, and Koll}]{rose_dependence_2014}
Rose, B. E.~J., K.~C. Armour, D.~S. Battisti, N.~Feldl, and D.~D.~B. Koll, 2014: The dependence of transient climate sensitivity and radiative feedbacks on the spatial pattern of ocean heat uptake. \textit{Geophysical Research Letters}, \textbf{41~(3)}, 1071--1078, \doi{10.1002/2013GL058955}.

\bibitem[{Rose and Marshall(2009)Rose, and Marshall}]{rose_ocean_2009}
Rose, B. E.~J., and J.~Marshall, 2009: Ocean {Heat} {Transport}, {Sea} {Ice}, and {Multiple} {Climate} {States}: {Insights} from {Energy} {Balance} {Models}. \textit{Journal of the Atmospheric Sciences}, \textbf{66~(9)}, 2828--2843, \doi{10.1175/2009JAS3039.1}.

\bibitem[{Rugenstein et~al.(2016)Rugenstein, Caldeira,, and Knutti}]{rugenstein_dependence_2016}
Rugenstein, M. A.~A., K.~Caldeira, and R.~Knutti, 2016: Dependence of global radiative feedbacks on evolving patterns of surface heat fluxes. \textit{Geophysical Research Letters}, \textbf{43~(18)}, 9877--9885, \doi{10.1002/2016GL070907}.

\bibitem[{Santer et~al.(1990)Santer, Wigley, {Schlesinger, Michael E.},, and {Mitchell, John F.B.}}]{santer_developing_1990}
Santer, B.~D., T.~M. Wigley, {Schlesinger, Michael E.}, and {Mitchell, John F.B.}, 1990: Developing climate scenarios from equilibrium {GCM} results. Tech. rep., MPI Report Number 47, Hamburg.

\bibitem[{Schaller et~al.(2014)Schaller, Sedláček,, and Knutti}]{schaller_asymmetry_2014}
Schaller, N., J.~Sedláček, and R.~Knutti, 2014: The asymmetry of the climate system's response to solar forcing changes and its implications for geoengineering scenarios. \textit{Journal of Geophysical Research: Atmospheres}, \textbf{119~(9)}, 5171--5184, \doi{10.1002/2013JD021258}.

\bibitem[{Sellers(1969)}]{sellers_global_1969}
Sellers, W.~D., 1969: A {Global} {Climatic} {Model} {Based} on the {Energy} {Balance} of the {Earth}-{Atmosphere} {System}. \textit{Journal of Applied Meteorology}, \textbf{8~(3)}, 392--400, \doi{10.1175/1520-0450(1969)008<0392:AGCMBO>2.0.CO;2}.

\bibitem[{Serreze and Barry(2011)Serreze, and Barry}]{serreze_processes_2011}
Serreze, M.~C., and R.~G. Barry, 2011: Processes and impacts of {Arctic} amplification: {A} research synthesis. \textit{Global and Planetary Change}, \textbf{77~(1-2)}, 85--96, \doi{10.1016/j.gloplacha.2011.03.004}.

\bibitem[{Seshadri(2017)}]{seshadri_origin_2017}
Seshadri, A.~K., 2017: Origin of path independence between cumulative {CO2} emissions and global warming. \textit{Climate Dynamics}, \textbf{49~(9-10)}, 3383--3401, \doi{10.1007/s00382-016-3519-3}.

\bibitem[{Solomon et~al.(2009)Solomon, Plattner, Knutti,, and Friedlingstein}]{solomon_irreversible_2009}
Solomon, S., G.-K. Plattner, R.~Knutti, and P.~Friedlingstein, 2009: Irreversible climate change due to carbon dioxide emissions. \textit{Proceedings of the National Academy of Sciences}, \textbf{106~(6)}, 1704--1709, \doi{10.1073/pnas.0812721106}.

\bibitem[{Stewart et~al.(2020)Stewart, Turner,, and Matthews}]{stewart_climate_2020}
Stewart, B.~M., S.~E. Turner, and H.~D. Matthews, 2020: Climate change impacts on potential future ranges of non-human primate species. \textit{Climatic Change}, \textbf{162~(4)}, 2301--2318, \doi{10.1007/s10584-020-02776-5}.

\bibitem[{Sutton et~al.(2007)Sutton, Dong,, and Gregory}]{sutton_landsea_2007}
Sutton, R.~T., B.~Dong, and J.~M. Gregory, 2007: Land/sea warming ratio in response to climate change: {IPCC} {AR4} model results and comparison with observations. \textit{Geophysical Research Letters}, \textbf{34~(2)}, 2006GL028\,164, \doi{10.1029/2006GL028164}.

\bibitem[{Södergren et~al.(2018)Södergren, McDonald,, and Bodeker}]{sodergren_energy_2018}
Södergren, A.~H., A.~J. McDonald, and G.~E. Bodeker, 2018: An energy balance model exploration of the impacts of interactions between surface albedo, cloud cover and water vapor on polar amplification. \textit{Climate Dynamics}, \textbf{51~(5-6)}, 1639--1658, \doi{10.1007/s00382-017-3974-5}.

\bibitem[{Tebaldi and Arblaster(2014)Tebaldi, and Arblaster}]{tebaldi_pattern_2014}
Tebaldi, C., and J.~M. Arblaster, 2014: Pattern scaling: {Its} strengths and limitations, and an update on the latest model simulations. \textit{Climatic Change}, \textbf{122~(3)}, 459--471, \doi{10.1007/s10584-013-1032-9}.

\bibitem[{Watson‐Parris et~al.(2022)}]{watsonparris_climatebench_2022}
Watson‐Parris, D., and Coauthors, 2022: {ClimateBench} v1.0: {A} {Benchmark} for {Data}‐{Driven} {Climate} {Projections}. \textit{Journal of Advances in Modeling Earth Systems}, \textbf{14~(10)}, e2021MS002\,954, \doi{10.1029/2021MS002954}.

\bibitem[{Wells et~al.(2023)Wells, Jackson, Maycock,, and Forster}]{wells_understanding_2023}
Wells, C.~D., L.~S. Jackson, A.~C. Maycock, and P.~M. Forster, 2023: Understanding pattern scaling errors across a range of emissions pathways. \textit{Earth System Dynamics}, \textbf{14~(4)}, 817--834, \doi{10.5194/esd-14-817-2023}.

\bibitem[{Womack et~al.(2024)Womack, Giani, Eastham,, and Selin}]{womack2024rapid}
Womack, C., P.~Giani, S.~D. Eastham, and N.~E. Selin, 2024: Rapid emulation of spatially resolved temperature response to effective radiative forcing. \textit{Authorea Preprints}.

\bibitem[{Zickfeld et~al.(2012)Zickfeld, Arora,, and Gillett}]{zickfeld_is_2012}
Zickfeld, K., V.~K. Arora, and N.~P. Gillett, 2012: Is the climate response to {CO} $_{\textrm{2}}$ emissions path dependent? \textit{Geophysical Research Letters}, \textbf{39~(5)}, 2011GL050\,205, \doi{10.1029/2011GL050205}.

\end{thebibliography}

\end{document}